\documentclass{pasj00}
\draft

\usepackage{color}
\usepackage{lscape}

\begin{document}
\SetRunningHead{M. Kawaharada et al.}{Constraints on Soft and Hard Excess Emission of Abell~2199}
\Received{2009/07/09}
\Accepted{2009/11/26}

\title{Suzaku Constraints on the Soft and Hard Excess Emissions from Abell~2199}

%


 \author{%
   Madoka \textsc{kawaharada,}\altaffilmark{1}
   Kazuo \textsc{makishima,}\altaffilmark{2,1}
   Takao \textsc{kitaguchi,}\altaffilmark{2}
   Sho \textsc{okuyama,}\altaffilmark{2}
   Kazuhiro \textsc{nakazawa,}\altaffilmark{2}
   and
   Yasushi \textsc{fukazawa}\altaffilmark{3}
}
 \altaffiltext{1}{Cosmic Radiation Laboratory, RIKEN, 2-1 Hirosawa, Wako, Saitama 351--0198.}
 \email{E-mail : kawahard@crab.riken.jp}
 \altaffiltext{2}{Department of Physics, The University of Tokyo, 7-3-1 Hongo,  Bunkyo-ku,  Tokyo 113--0033.}
 \altaffiltext{3}{Department of Physical Science, Hiroshima University, 1-3-1 Kagamiyama, Higashi-hiroshima 739-8526.}

\KeyWords{galaxies: clusters: individual (Abell~2199) --- X-rays: galaxies: clus
ters} 

\maketitle

\begin{abstract}
The nearby ($z=0.03015$) cluster of galaxies Abell~2199 was observed by 
Suzaku in X-rays, with five pointings for $\sim 20$ ks each. 
From the XIS data, the temperature and metal abundance
profiles were derived out to $\sim 700$ kpc (0.4 times virial radius).   
Both these quantities decrease gradually from the 
center to peripheries by a factor of $\sim 2$, while the oxygen 
abundance tends to be flat. The temperature within $12'$ ($\sim 430$ kpc) is $\sim 4$ keV, 
and the 0.5--10 keV X-ray luminosity integrated up to $30'$
is $(2.9 \pm 0.1) \times 10^{44}$ erg s$^{-1}$, 
in agreement with previous XMM-Newton measurements. 
Above this thermal emission, no significant excess was found either
in the XIS range below $\sim 1$ keV, or in the HXD-PIN range above $\sim 15$ keV.
The 90\%-confidence upper limit on the emission measure of an assumed 0.2 keV warm gas is
(3.7--7.5) $\times 10^{62}$ cm$^{-3}$ arcmin$^{-2}$,  which is
3.7--7.6 times tighter than the detection reported with XMM-Newton. 
The 90\%-confidence upper limit on the 20--80 keV luminosity of any power law 
component is $1.8 \times 10^{43}$ erg s$^{-1}$, assuming a photon
index of 2.0. Although this upper limit does not reject the possible 2.1$\sigma$
detection by the BeppoSAX PDS, it is a factor of 2.1 tighter than 
that of the PDS if both are considered upper limits. 
The non-detection of the hard excess can be reconciled
with the upper limit on diffuse radio emission, 
without invoking the very low magnetic fields ($< 0.073 \mu$G) which 
were suggested previously.
\end{abstract}

\section{Introduction}
The intra-cluster medium (ICM) in clusters of galaxies, i.e,  
hot thermal plasmas in collisional ionization equlibria
confined within their gravitational potential,
constitutes the most dominant form of cosmic baryons that
has ever been detected. 
Although the ICM radiates predominantly optically-thin thermal
X-ray emission, some clusters have been reported to exhibit
excess signals above the thermal emission at the lowest or highest 
energy ends of their X-ray spectra.
One interpretation of the soft excess is emission from Warm Hot Intergalactic 
Medium (WHIM), which has been predicted from cosmological simulations (e.g. \cite{cen-1999}) and is expected to solve the so-called missing baryon problem 
(e.g \cite{fukugita-1998}). Another interpretation of the spectral soft 
and/or hard excess is non-thermal emission from accelerated particles 
in galaxy clusters (e.g. \cite{lieu-1999}; \cite{fusco-1999}):
a power-low shaped spectrum can exceed the dominant thermal emission at
both sufficiently high and low energies. 
This provides a possibility that galaxy
clusters are giant accelerators in the universe. Thus, the search for
soft and hard excess signals in clusters forms a very important
research subject. However, the instrumental sensitivity in these photon energies
has been insufficient, and hence the existence of these excess 
components has remained controversial 
(e.g. \cite{bregman-2006}; \cite{nevalainen-2007}; \cite{rossetti-2004}; \cite{fusco-2007}).

Among those galaxy clusters from which the excess emission has been reported, 
Abell~2199 is particularly interesting, because it has been 
suspected to exhibit both the soft and hard excess components. 
In fact, Kaastra et al. (1999) 
reported the detection of both, based on the 
BeppoSAX, Extreme Ultraviolet Explorer, and ROSAT observations. 
The two excess components were simultaneously expressed by a single
power law with a photon index $\sim 1.8$ which is superposed on the 
thermal emission spectrum, and interpreted as inverse Compton
(IC) photons produced when the cosmic microwave background (CMB) 
photons are scattered up by relativistic electrons in the cluster. 
However, as argued by \citet{kempner-2000}, the magnetic field 
in Abell~2199 would have to be unusually weak ($< 0.073\; \mu$G) in order 
for the relativistic electrons not to produce synchrotron emission
beyond measured upper limits of diffuse radio emission. 
As an alternative interpretation, \citet{kempner-2000} proposed non-thermal bremsstrahlung by 
supra-thermal electrons which are being accelerated in the cluster. 

In the era of XMM-Newton, the possibility of thermal emission from a 
warm gas also became another popular idea to explain the soft excess in
several clusters. Kaastra et al. (2004) reported a redshifted 
O~VII emission line from a 0.2 keV warm gas in Abell~2199, 
along with Coma, Abell~1795, Sersic~159-03, MKW~3s, Abell~2052, 
and Abell~3112, and hence ascribed their soft excess to emission from WHIM.
At the same time, Chandra has made progress on the ICM physics of Abell~2199
especially in the central region, such as an asymmetric temperature distribution
in the direction perpendicular to the jets within $30''$ of the center
(Kawano et al. 2003), and a weak isothermal shock associated with the central 
active galactic nucleus (Sanders \& Fabian 2006).

In order to examine the soft and hard excess phenomena of Abell~2199 with
a better sensitivity, we observed the object with Suzaku
(\cite{mitsuda-2007}). 
The excellent low-energy capability of XIS-BI 
(back illuminated CCD, \cite{koyama-2007}) 
and the low background of the silicon PIN detectors in the HXD 
(\cite{takahashi-2007}; \cite{kokubun-2007}) have provided the
best data in the soft and hard X-ray bands, respectively. 
As a result, we have obtained tight upper limits on both excess components,
and strengthened the dominance of the thermal emission from this cluster.


In the present paper, 
quantities quoted from the literature are rescaled to 
the Hubble constant of
$H_0 = 71\; \rm{km\; s^{-1}\; Mpc^{-1}}$, and
errors are given at 90 \% confidence range unless otherwise stated.

\section{Observation and Data Preparation 
\label{section:obs}}
\subsection{\it Observation
\label{subsection:obs:obs}}
\begin{figure}
  \begin{center}
   \includegraphics[width=0.33\textwidth,angle=0,clip]{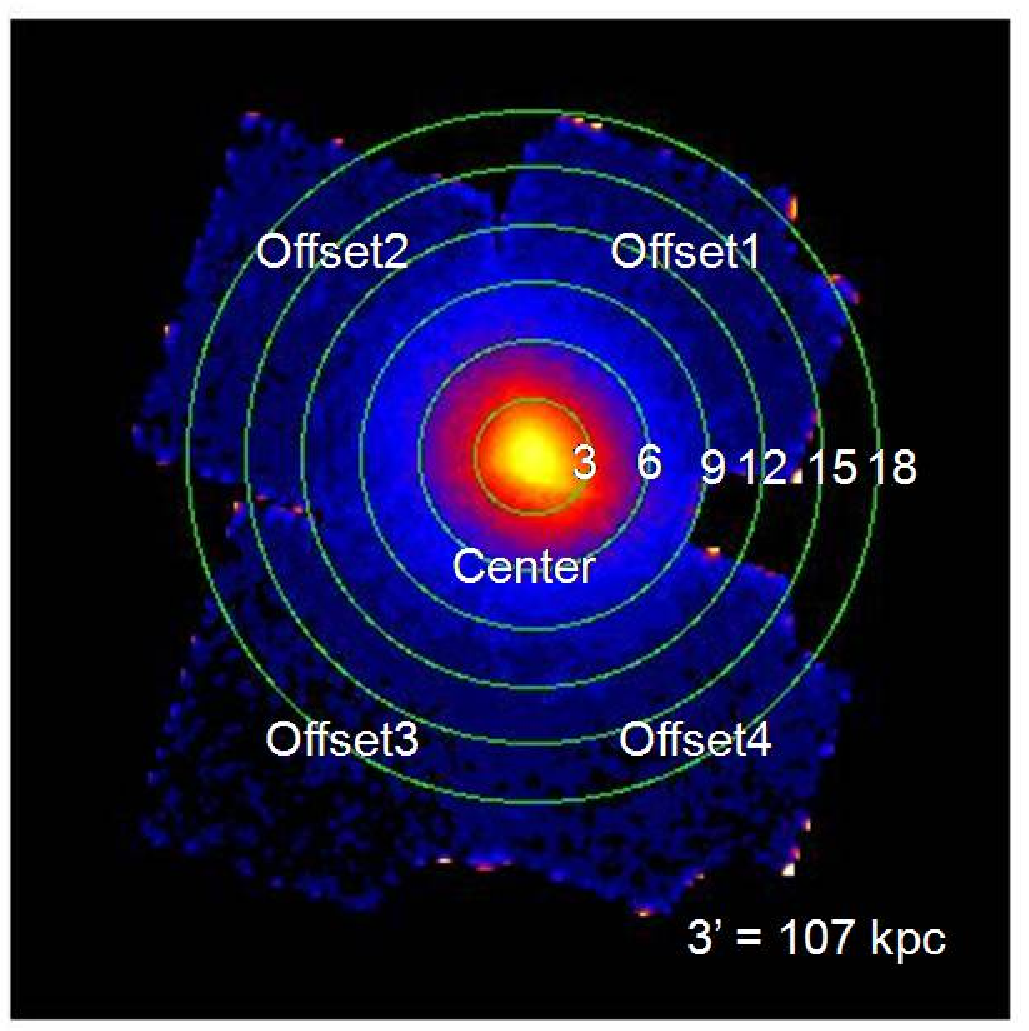} 
  \end{center}
  \caption{A 0.5--10 keV mosaic XIS~0 image from the Center, Offset1, Offset2, Offset3, 
and Offset4 observations, smoothed 
with a gaussian of $\sigma = 10''$ and corrected for vignetting.
Boundaries of annular regions used for the spectral analysis are shown in green circles.
See the electronic version of the paper for a colour figure. 
}
\label{fig:mosaic}
\end{figure}

As shown in  figure~\ref{fig:mosaic} and table~\ref{table:observation}, 
we observed Abell 2199 with Suzaku in the beginning of
October 2006. The observation was conducted in 5 separate pointings
(``Center'' and ``Offset1'' through ``Offset4''), for
$\sim 20$ ks each, with the XIS field of view (FOV) partially overlapping. 
Throughout the 5 pointings, all the XIS sensors 
were in normal clock mode without window or burst options.
These observations became one of the first ones 
in which spaced-row charge injection (SCI) 
was applied to the XIS detectors (Nakajima et al. 2008). 
In the present paper, we use data of XIS0, XIS1, XIS3
and the HXD-PIN. Although XIS2 was still operational, we do not use
its data because a database of non X-ray background (NXB) for SCI-ON XIS2 is too small, 
which is available only until the anomaly in XIS~2 on 2009 November 9\footnote{http://www.astro.isas.ac.jp/suzaku/doc/suzakumemo-2007-08.pdf}.

Between the Offset2 and Offset4 observations, operations
of changing the HXD-PIN bias voltage were carried out in response to 
flare-like events in a PIN diode\footnote{http://www.astro.isas.jaxa.jp/suzaku/log/hxd/} (W10P0). 
As summarized in table~\ref{table:hv-ope},
data of all the 64 PINs were available throughout the Center and Offset1
pointings, and for a part of Offset2 and 
Offset4. In contrast, only one fourth of the PIN diodes were 
operational in Offset3. Therefore, 
we do not analyze the Offset3 HXD-PIN data in the present paper. For the
other observations, we exclude the data of the four PIN diodes in W10 unit, 
and use the remaining 60 channels.  Backgrounds and responses excluding
the four PIN diodes were prepared with a help of the HXD team.

\subsection{\it XIS data reduction 
\label{subsection:obs:xis}}
The present paper utilizes the XIS and HXD-PIN data, prepared through
version 2.0 pipe-line data processing, and obtained from the Suzaku 
ftp site\footnote{ftp://ftp.darts.isas.jaxa.jp/pub/suzaku/ver2.0/}.
We re-processed unscreened XIS
event files using Suzaku software version 11 in HEAsoft 6.6,
together with the calibration datebase (CALDB) released on 2009 January
9. We applied \verb+xiscoord+, \verb+xisputpixelquality+, 
\verb+xispi+, and \verb+xistime+ in this order, and performed the standard
event screening. Bad pixels were rejected with \verb+cleansis+, 
employing the option of chipcol=SEGMENT.
We selected events with GRADE 0,2,3,4, and 6. Good time intervals (GTI)
were detemined by criteria as
 \verb+"SAA_HXD==0+   \verb+&&+ \verb+T_SAA_HXD+\verb+>436 && COR>6 && ELV>5+ \verb+&& DYE_ELV>20 && AOCU_HK_CNT3_NML_P==1 && Sn_DTRATE<3 && ANG_DIST<1.5"+. Details of the processing and
screening are the same as those described in the Suzaku Data Reduction Guide.\footnote{http://ftools.gsfc.nasa.gov/docs/suzaku/analysis/abc/}

The redistribution matrix files (RMFs) of the XIS were produced by 
\verb+xisrmfgen+, and auxillary responce files (ARFs) 
by \verb+xissimarfgen+ (\cite{ishisaki-2007}). 
As an input to the ARF generator, we prepared an X-ray surface brightness 
profile of Abell~2199 using a double-$\beta$ model (sum of two $\beta$ models; \cite{king-1962})
of which the parameters were determined through a least chi-square fit to
the background subtracted and vignetting corrected 
0.5--10 keV XMM-Newton MOS1 image of the Abell~2199 cluster. 
The employed parameters are 
($r_{\rm c}$, $\beta$)$=$($10.3 \pm 0.5$ kpc, $0.650 \pm 0.008$) for the narrower component
and ($51.0 \pm 0.8$ kpc, $0.531 \pm 0.002$) for the wider component, 
where $r_{\rm c}$ is the core radius. 
Absorption below 2 keV, caused by a carbon-dominated contamination material on the XIS 
optical blocking filters, is included in the ARFs. 
There, differences of the contaminant thickness among the XIS sensors are taken into
account, along with its radial dependences and secular changes (\cite{koyama-2007}).

Non X-ray background (NXB) of each XIS sensor was created 
using \verb+xisnxbgen+, 
which sorts spectra of night earth observations according to
the geomagnetic cut-off-rigidity (COR) and makes an averaged  spectrum
weighted by residence times for which Suzaku stayed in each COR interval
(\cite{tawa-2008}). 
We assumed systematic errors (90\% confidence range) of 6.0\% and 12.5\%
for the NXB from XIS-FI (front illuminated CCD,  \cite{koyama-2007}) and XIS-BI 
(\cite{tawa-2008}; table 5 (c)), respectively, and added them in quadrature to the
corrensponding statistical errors. The X-ray background was estimated as described 
in subsection~\ref{subsection:bgd:xis}.




\subsection{\it HXD-PIN data reduction 
\label{subsection:obs:pin}}
Unscreened event files of HXD-PIN were re-processed by
\verb+hxdtime+, \verb+hxdpi+, and \verb+hxdgrade+ in this order. 
Using \verb+hxdgtigen+, we rejected the time intervals when ``FIFO Full'', 
``BFSH'', or ``TLMRJCT'' happened, and further narrowed the
GTI by imposing criteria of \verb+"SAA_HXD==0 &&+ \verb+T_SAA_HXD>500 &&+ \verb+TN_SAA_HXD>180 &&+ \verb+COR>6 &&+ \verb+ELV>5 &&+ \verb+AOCU_HK_CNT3_NML_P==1 &&+ \verb+HXD_DTRATE<3 &&+ \verb+ANG_DIST<1.5 &&+ \verb+HXD_HV_Wn_CAL>700 &&+ \verb+HXD_HV_Tn_CAL>700"+.   

For the Center, Offset1, Offset2, and Offset4 pointings,
we made special HXD-PIN RMFs which exclude contributions of the four PINs
in W10 unit. This was constructed by summing RMFs from the remaining 60 PIN diodes.
Since \verb+hxdarfgen+ did not support extended sources yet, 
we created PIN ARFs for each  $1' \times 1'$ pixel by \verb+hxdarfgen+, 
assuming the same X-ray surface brightness as used
for XIS ARFs (subsection~\ref{subsection:obs:xis}) . 
The ARFs were then averaged by weighting with counts contained in the corresponding pixels. 

As the HXD-PIN NXB,  the ``tuned'' PIN NXB files published by the HXD 
team\footnote{http://www.astro.isas.jaxa.jp/suzaku/analysis/hxd/pinnxb/tuned/}
were used (\cite{fukazawa-2009}). 
In the present paper, we used 2.3\% as systematic error (90\% confidence range) of the PIN NXB.
Details of its derivation is described in
Appendix~\ref{appen:nxb-sys}.
The PIN CXB was estimated from an observation of a nearby region
in the same manner as the XIS CXB (subsection~\ref{subsection:bgd:pin}).  


\section{Estimation of X-ray Backgrounds
\label{section:bgd}}

\subsection{\it XIS X-ray background
\label{subsection:bgd:xis}}
To analyze the XIS data of an extended object in general, we must
subtract the Cosmic X-ray background (CXB) and Galactic foreground
emission (GFE).
Since the emission from Abell~2199 itself hampers their direct estimation 
using our own data, we used 73.6 ks 
Suzaku observation of a nearby blank sky region named 
``High Latitude Diffuse A'' 
(Hereafter HLD; Observation ID = 500027010), which is $4^{\circ}.03$ 
away from Abell~2199.
The difference of their Galactic 
latitudes is only $0^{\circ}.7$, and hence the hydrogen column 
densities (\cite{dickey-1990}, weighted average over $1^{\circ}$ cone radius) 
toward Abell~2199 ($0.86 \times 10^{20}$ cm$^{-2}$) 
and HLD ($1.02 \times 10^{20}$ cm$^{-2}$) are similar. 

After applying the standard processing described in subsection~\ref{subsection:obs:xis}
to the unscreened XIS data of HLD, we removed circular regions ($1'$ in radius) around 
six point sources that are visible in the 0.5--10.0 keV XIS images, 
and extracted spectra of the remaining region from the four XIS detectors.
The corresponding detection limit for the point sources  is 
$\sim 6 \times 10^{-14}$ erg s$^{-1}$ cm$^{-2}$ in 2--10 keV.
Then, the NXB-subtracted XIS-BI and XIS-FI 
(averaged over the three XIS-FI detectors) spectra in energy ranges of 
0.3--5.5 keV and 0.5--5.5 keV, respectively,
were fitted simultaneously with a CXB+GFE model (see below) using XSPEC12 
version 12.4.0ad,
incorporating the XIS response for a source of uniform brightness. 
In the analyses described below, metal abundances refer to
Anders \& Grevesse (1989), and the photoelectric absorption 
cross-sections to Balucinska-Church \& McCammon (1992) with a new 
He cross section (\cite{yan-1998}).  

As the CXB model, we used a power law with a fixed photon index of 1.41
(\cite{kushino-2002}) and a free normalization.
To represent the GFE, we employed model 1 of \citet{henley-2008}, 
which consists of three {\it apec} components
(\cite{smith-2001}): a non-absorbed 0.08 keV component, an absorbed 
0.11 keV one, and an absorbed 0.27 keV one, representing the emission from
Local bubble (LB), a cool halo, and a hot halo, respectively. The three  {\it apec}
temperatures were all fixed. 
The hydrogen column density was fixed to the Galatic value of $1.02 \times 10^{20}$~cm$^{-2}$,
and the metal abundance and redshift were also fixed to one solar and zero, 
respectively. Relative normalizations of the two halo components were 
tied as cool halo : hot halo $= 1 : 0.24$ (\cite{henley-2008}).
Thus, the GFE model has only two free parameters; normalizations of the 
LB and halo. 
As summarized in table~\ref{table:fit-bgd}, this CXB+GFE model
reproduced the XIS spectra from HLD successfully. 
The 2.0--10.0 keV surface brightness of the CXB component is
$(6.34^{+0.13}_{-0.12} \pm 0.63) \times 10^{-8}$ erg s$^{-1}$ cm$^{-2}$ sr$^{-1}$ 
(90\% statistical and systematic errors),
where the systematic error, 10\%, refers to Appendix 1 of \cite{nakazawa-2009}. 
Using ASCA,  \citet{kushino-2002} derived an absolute 2.0--10.0 keV CXB surface brightness
to be $(6.38 \pm 0.07 \pm 1.05) \times 10^{-8}$ erg s$^{-1}$ cm$^{-2}$ sr$^{-1}$ 
(90\% statistical and systematic errors). 
This value becomes
$(4.2$ -- $6.0) \times 10^{-8}$ erg s$^{-1}$ cm$^{-2}$ sr$^{-1}$ 
when the surface brightness is integrated to our detection limit of $6 \times 10^{-14}$ erg s$^{-1}$ cm$^{-2}$ 
(equation 6 of \cite{kushino-2002}).
Thus, the CXB level obtained from the HLD observation is consistent
with the ASCA result. 

\subsection{\it  HXD-PIN X-ray background
\label{subsection:bgd:pin}}

We also analyzed the HXD-PIN data of the HLD observation which
were processed by the same standard procedure as described 
in subsection~\ref{subsection:obs:pin}.
Ignoring the GFE which is negligible at $\gtrsim 3$ keV, 
we fitted the NXB-subtracted $12$--$40$ keV HXD-PIN spectrum 
with a wide-band CXB model. 
Namely, following the HEAO-1 result (\cite{boldt-1987}), 
we expressed the CXB spectral surface brightness as
\begin{equation}
S_{\rm X}(E) = S_0\; \left(\frac{E}{3\; {\rm keV}}\right)^{-0.29} {\rm exp}\left(\frac{-E}{40\; {\rm keV}}\right)\ {\rm erg}\; {\rm cm}^{-2}\; {\rm s}^{-1}\; {\rm sr}^{-1}\; {\rm keV}^{-1}, \label{eqn:pin-cxb}
\end{equation}
where $S_0$, a normalization, is a free parameter in the fit.
When reproducing the CXB signals in HXD-PIN, equation~\ref{eqn:pin-cxb} 
was corrected for the 13\% systematic 
difference of cross-normalization between the XIS and 
PIN\footnote{http://www.astro.isas.jaxa.jp/suzaku/doc/suzakumemo/suzakumemo-2007-11.pdf}. 
The result we obtained, $S_0 = (9.7 \pm 2.0) \times 10^{-9}$,  
is consistent with the HEAO result of $S_0 = 9.0 \times 10^{-9}$, which can in turn 
reproduce, within 1\%, blank-sky PIN spectra which have higher statistics (\cite{fukazawa-2009}). 
The contribution of the six point sources to the HXD-PIN spectrum is
an order of magnitude lower than the NXB-subtracted signal in 12--40 keV. 
Therefore, the contribution of point sources affects the CXB
level only by $\Delta S_0 = 0.1 \times 10^{-9}$, 
which is much smaller than the statistical error.

\begin{figure}
  \begin{center}
   \includegraphics[width=0.25\textwidth,angle=-90,clip]{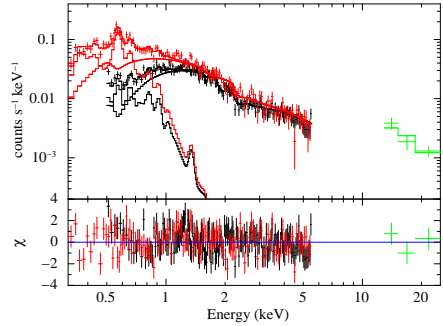} 
  \end{center}
  \caption{NXB-subtracted XIS-FI (black; 0.5--5.5 keV), XIS-BI (red; 0.3-5.5 keV), 
and HXD-PIN (green; 12-30 keV) spectra of the HLD observation, fitted with the X-ray 
background model described in the text. The CXB is modeled by 
a power-law in the XIS band, and equation~\ref{eqn:pin-cxb} in the
HXD-PIN region, with their surface brightness constrained to match over 6.0--12.0 keV. 
A model for the GFE is included in the XIS range. 
}
\label{fig:hld-spec}
\end{figure}

Although we have thus adopted the two separate models to estimate the
CXB in the XIS and HXD-PIN, the derived two surface brightness results
agree within 1\% in the connecting energy range of 
6.0--12.0 keV. Therefore, our two modelings of the CXB are consistent.
If we constrain the two CXB models so that the 6.0--12.0 keV surface brightness becomes the same,
and fit the XIS and HXD-PIN spectra simultaneously,
we can constrain the CXB above 10 keV better with
$S_0 = (9.7 \pm 0.2) \times 10^{-9}$.  This case is shown in figure~\ref{fig:hld-spec}.


\subsection{\it  Field to field difference of X-ray background
\label{subsection:bgd:sys}}
In the following spectral analysis, we fit the Abell~2199 data
simultaneously with the HLD data, in order to determine the X-ray
background. However, the true X-ray background of the Abell~2199 field
may be different from that of the HLD field. 
In order to examine the GFE brightness for possible differences between
them, we employed X-ray count rates 
of these two fields from the ROSAT All-Sky Survey (RASS) 
which are available via the NASA's HEASARC 
website\footnote{http://heasarc.gsfc.nasa.gov/cgi-bin/Tools/xraybg/xraybg.pl}.
Since the RASS data right on the Abell~2199 field is contaminated, of course,
with the cluster emission, we obtained a count rate averaged over an 
annular region of which the inner radius is $1^{\circ}.0$ ($\sim 2.2$ Mpc) and 
the outer radius is $2^{\circ}.0$, centered on the cD galaxy NGC~6166.
Since the virial radius of Abell~2199 is estimated to be $\sim 1.7$ Mpc ($\sim 0^{\circ}.8$)
from the ICM temperature of 4 keV (\cite{evrard-1996}), this annular region
can be considered to be free from the ICM emission.

Table~\ref{table:rass-cnt} summarizes the RASS count rates of the
Abell~2199 annular region and the HLD field (circular region, $1^{\circ}.0$ in radius).
The 3/4 keV (0.47--1.21 keV) and 1.5 keV (0.76--2.04 keV) rates are 
the same between the two fields  within statistical errors. 
Thus, the HLD region is confirmed to be a good background estimator for our 
Abell~2199 observations in these energy bands.  
The 1/4 keV (0.12--0.284 keV) band rates differ by $\sim 20$\%, 
suggesting a spectral difference betweeen the Abell~2199 
and HLD fields in the softest energy band. 
However, the LB component, which dominates the spectrum in the 1/4 keV band, 
is poorly determined with the XIS spectra, and
the 20\% discrepancy is within the statistical error of
$\sim 40$\% associated with the LB normalization in the following spectral analysis
(\S~\ref{subsection:results:xis-ana}).

The CXB surface brightness is known to fluctuate from direction to
direction (\cite{ishisaki-1997}).  
In the case of Suzaku, this fluctuation is estimated 
to be 10\% and 18\% for the XIS (full FOV) and PIN, respectively 
(see Appendix of \cite{nakazawa-2009} for details). 
We take these systematic errors into account in the spectral
analyses described in section~\ref{section:results}.  



\section{Data Analysis and Results
\label{section:results}}

\subsection{\it Thermal emission
\label{subsection:results:xis-ana}}
In order to quantify the thermal emission of Abell~2199, 
we defined seven concentric annular regions as shown in 
figure~\ref{fig:mosaic}, each with a radial width of $3'$.
The center is chosen to be the X-ray emission centroid at 
($\alpha,\delta$)=($16^{\rm h} 28^{\rm m} 36.9^{\rm s}, +39^\circ 32' 53''$) 
in J2000.0 coordinates. This is $0'.4$ offset from the nucleus
of the cD galaxy, NGC~6166, but well inside its optical
extent of $\sim 2'$ in diameter.

\begin{figure}
  \begin{center}
   \includegraphics[width=0.25\textwidth,angle=-90,clip]{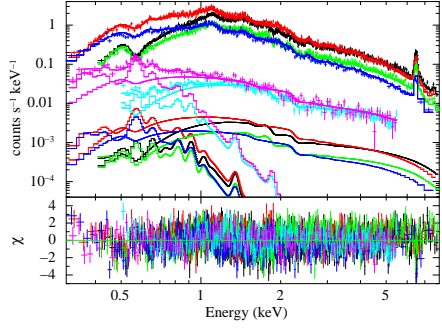} 
  \end{center}
  \caption{NXB-subtracted XIS spectra of Abell~2199 ($0'-3'$ region) and 
the HLD. Spectra from Center XIS-FI (black; 0.4--8.0 keV), Center XIS-BI (red; 0.3--8.0 keV), 
Offset1 XIS-FI (green; 0.4--8.0 keV), and Offset1 XIS-BI (blue; 0.3--8.0 keV) 
are fitted with a common \texttt{vapec} model and the X-ray background model. 
The XIS-FI (cyan; 0.5--5.5 keV) and XIS-BI (magenta; 0.3--5.5 keV) spectra
from the HLD field
are fitted simultaneously with the same X-ray background model. 
Further details are described in the text. 
}
\label{fig:xis_hld-spec}
\end{figure}

In analyzing a given annulus, we extracted XIS-FI (averaged over XIS 0 and XIS 3) 
and XIS-BI spectra from corresponding regions in the five pointings
(but excluding those which have no intersection),
and subtracted the NXB as described in subsection~\ref{subsection:obs:xis}.
For each annulus, the utilized observations are given in 
table~\ref{table:fit-reg}. Then, the spectra from the same annulus
(but different fields of view) were read into XSPEC together with 
the NXB-subtracted HLD spectra to constrain the X-ray background.
The annular spectra were fitted
simultaneously with a common {\it vapec} model 
with a free photoelectric absorption, together with the X-ray background model 
described in subsection~\ref{subsection:bgd:xis}, incorporating the
responses created for the Abell~2199 ICM emission (subsection~\ref{subsection:obs:xis})
and for a uniform X-ray source, respectively. 
All the parameters of the X-ray background model were tied  between
the annular and HLD spectra, except for the thickness of XIS contamination
which depends on the date of observation (\cite{koyama-2007}).
Among the different observations of Abell~2199, all the {\it vapec} parameters,
except the overall normalization, were constrained to be the same.
The cross normalization between XIS-FI and XIS-BI was set free. 
Metal abundances, $Z_{\rm metal}$, in the {\it vapec} model were tied 
in three groups as 
$Z_{\rm O}=Z_{\rm Ne}=Z_{\rm Mg}=Z_{\rm Al}$, 
$Z_{\rm Si}=Z_{\rm S}=Z_{\rm Ar}=Z_{\rm Ca}$, 
and $Z_{\rm Fe}=Z_{\rm Ni}$.
An example of this fitting in the $0'$--$3'$ region is shown 
in figure~\ref{fig:xis_hld-spec}.
We also changed the normalization of CXB model by $\pm 11$\% and
studied the effect of CXB fluctuation described in subsection~\ref{subsection:bgd:sys}.
Although the X-ray background model is determined separately for the seven individual
regions, the normalizations of CXB, LB, and Halo are consistent among the regions
within errors (see figure~\ref{fig:bgd-norm} in Appendix~\ref{appen:bgd-norm}).

The results of this analysis are summarized in figure~\ref{fig:vapec-prof} 
(for details, see table~\ref{table:fit-apec} in Appendix~\ref{appen:fit-result}).
The single-temperature {\it vapec} model thus reproduced the spectra 
successfully, and the CXB fluctuation did not affect the results
significantly. 
The temperature and iron abundance were determined 
out to $\sim 700 $ kpc, which is about 0.4 times the virial radius of 
a $4$ keV cluster. The temperature gradually decreases toward cluster
outskirts from $\sim 4$ keV to $\sim 3$ keV. The temperature within 
$12'$, $\sim 4$ keV, is consistent with that derived with ASCA 
(4.1 keV; \cite{fukazawa-2004}). 
The iron and silicon abundances also decrease outwards by a factor of $\sim 2$, 
while the oxygen abundance tends to be flat. The hydrogen column density obtained from the fit became
($2$--$3) \times 10^{20}$ cm$^{-2}$, which is significantly higher than the Galacitic value. 
We examine the cause of this effect in subsection~\ref{subsection:results:soft-excess}. 
The 0.5 -- 10 keV X-ray luminosity integrated out to $30'$
is $(2.9 \pm 0.1) \times 10^{44}$ erg s$^{-1}$, which is  consistent
with the result of XMM-Newton (\cite{snowden-2008}). 

Next, we attempted a two-temperature fit using a model expressed as {\it vapec}+{\it vapec}. 
Metal abundances of the two components were tied together.
As shown in figure~\ref{fig:2vapec-prof}a and figure~\ref{fig:2vapec-prof}f
(for details, see table~\ref{table:fit-2apec} in Appendix~\ref{appen:fit-result}),
the spectra in the innermost region ($0'$--$3'$) were reproduced better with 
the two-temperature model of 2~keV and 5~keV, while the improvement was 
not significant in the other regions.  This is reasonable, because \citet{johnstone-2002}
detected, with Chandra, a temperature drop from 5~keV to 2~keV toward the center within  
a central region of radius 100 kpc ($2'.8$). The iron and silicon abundance
profiles did not change significantly between the single-temperature 
and two-temperature models. However, the oxygen abundance in the inntermost region 
decreased and the profile became flatter. 
The high values of $N_{\rm H}$, found with the single-temperature analysis,
persisted; this effect cannot be regarded as an artifact caused by an
inappropriate temperature modeling. 


\subsection{\it Constraints on soft excess emission
\label{subsection:results:soft-excess}}
\begin{figure}
  \begin{center}
   \includegraphics[width=0.25\textwidth,angle=-90,clip]{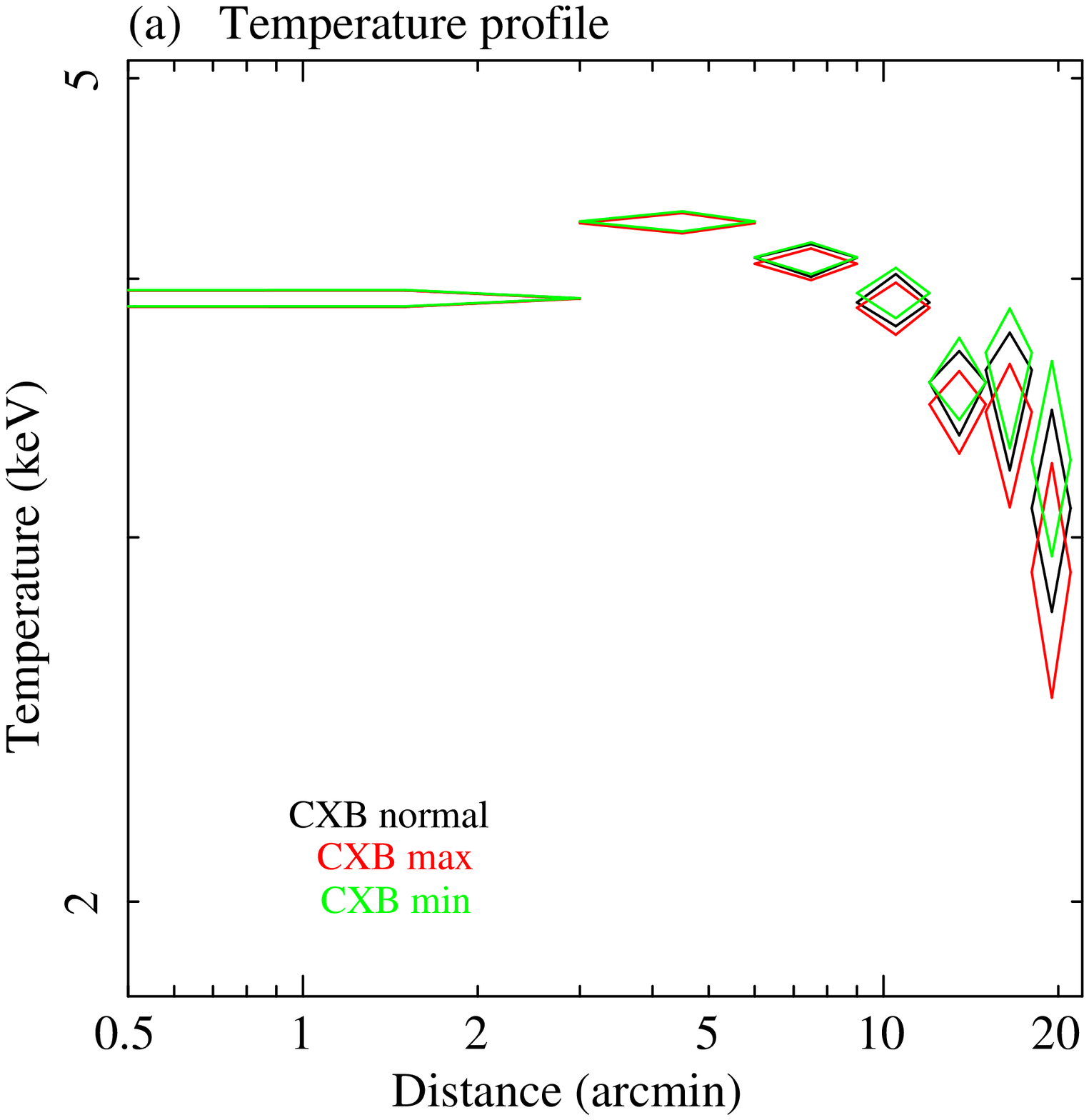} 
\hspace{0.3cm}
\vspace{0.3cm}
   \includegraphics[width=0.25\textwidth,angle=-90,clip]{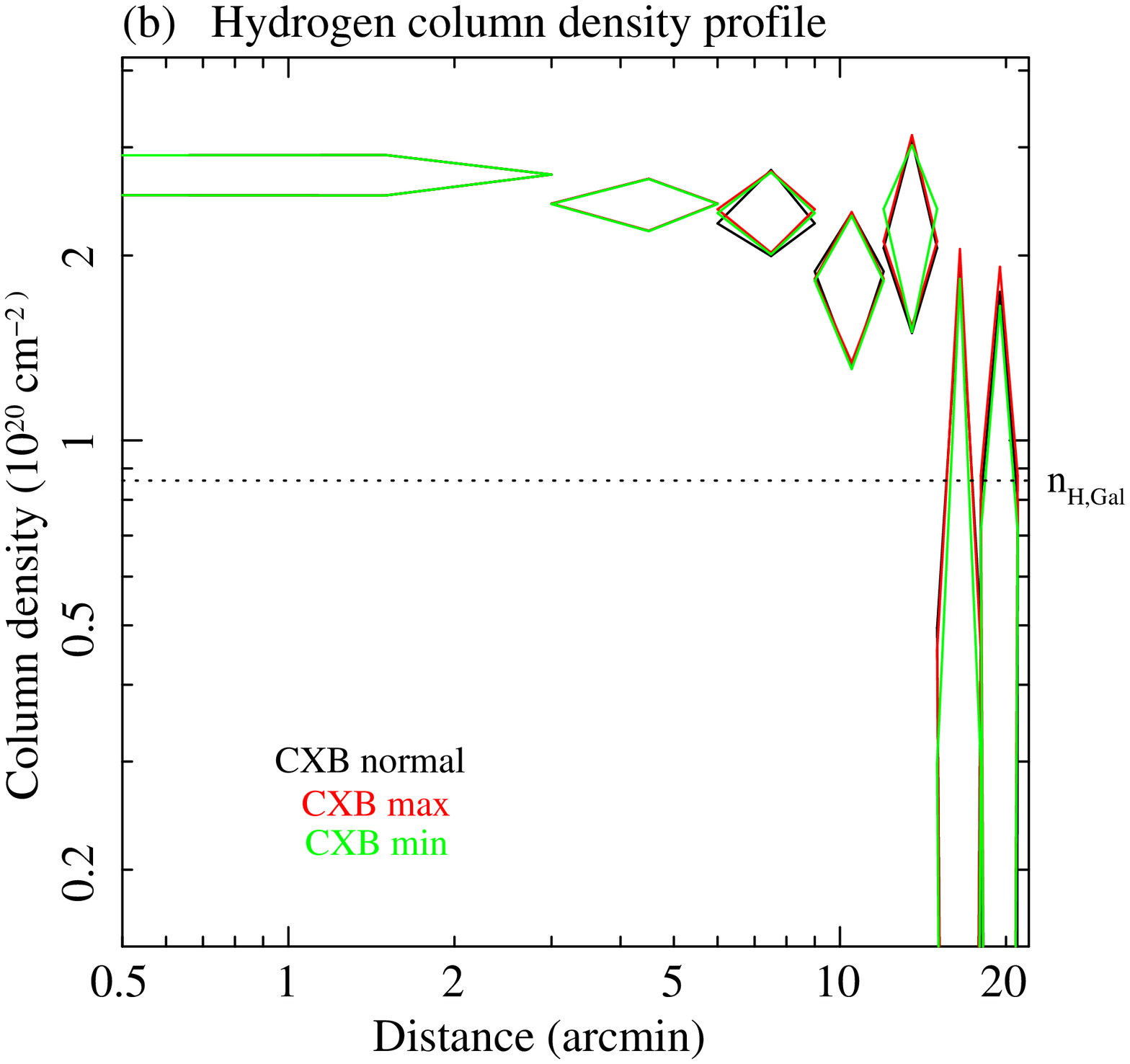} 
\hspace{0.3cm}
   \includegraphics[width=0.25\textwidth,angle=-90,clip]{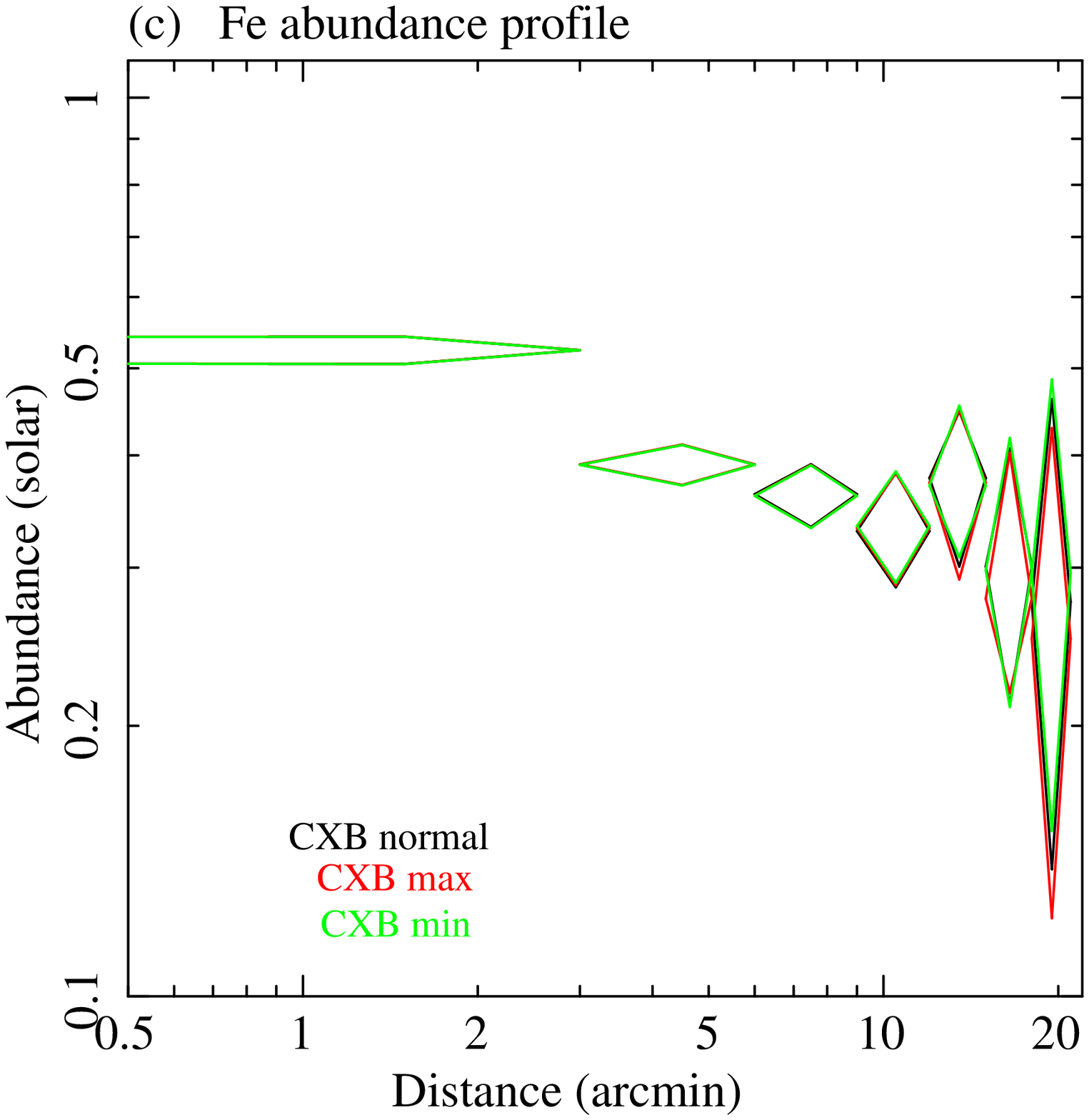} 
\hspace{0.3cm}
   \includegraphics[width=0.25\textwidth,angle=-90,clip]{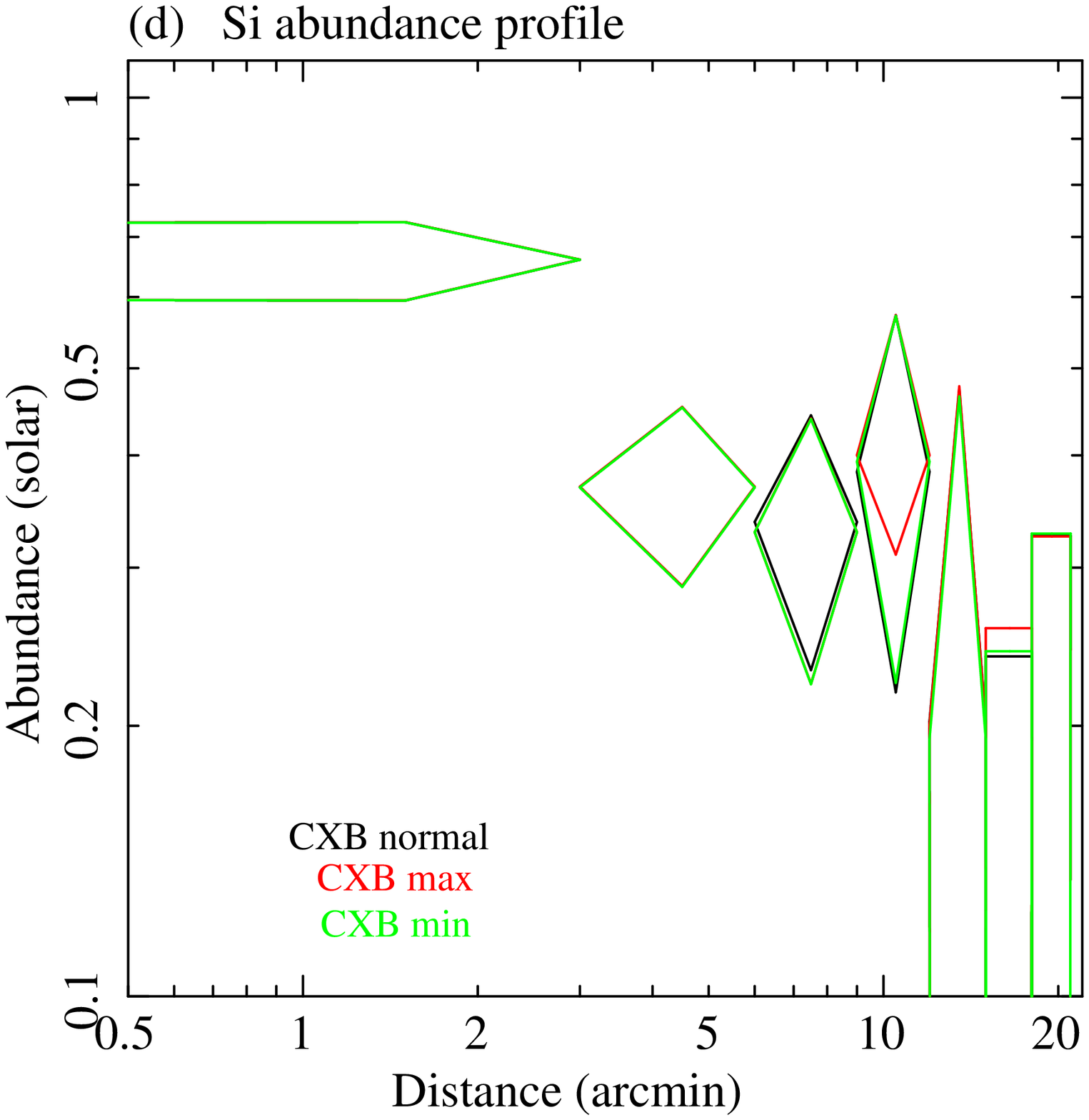} 
\hspace{0.3cm}
   \includegraphics[width=0.25\textwidth,angle=-90,clip]{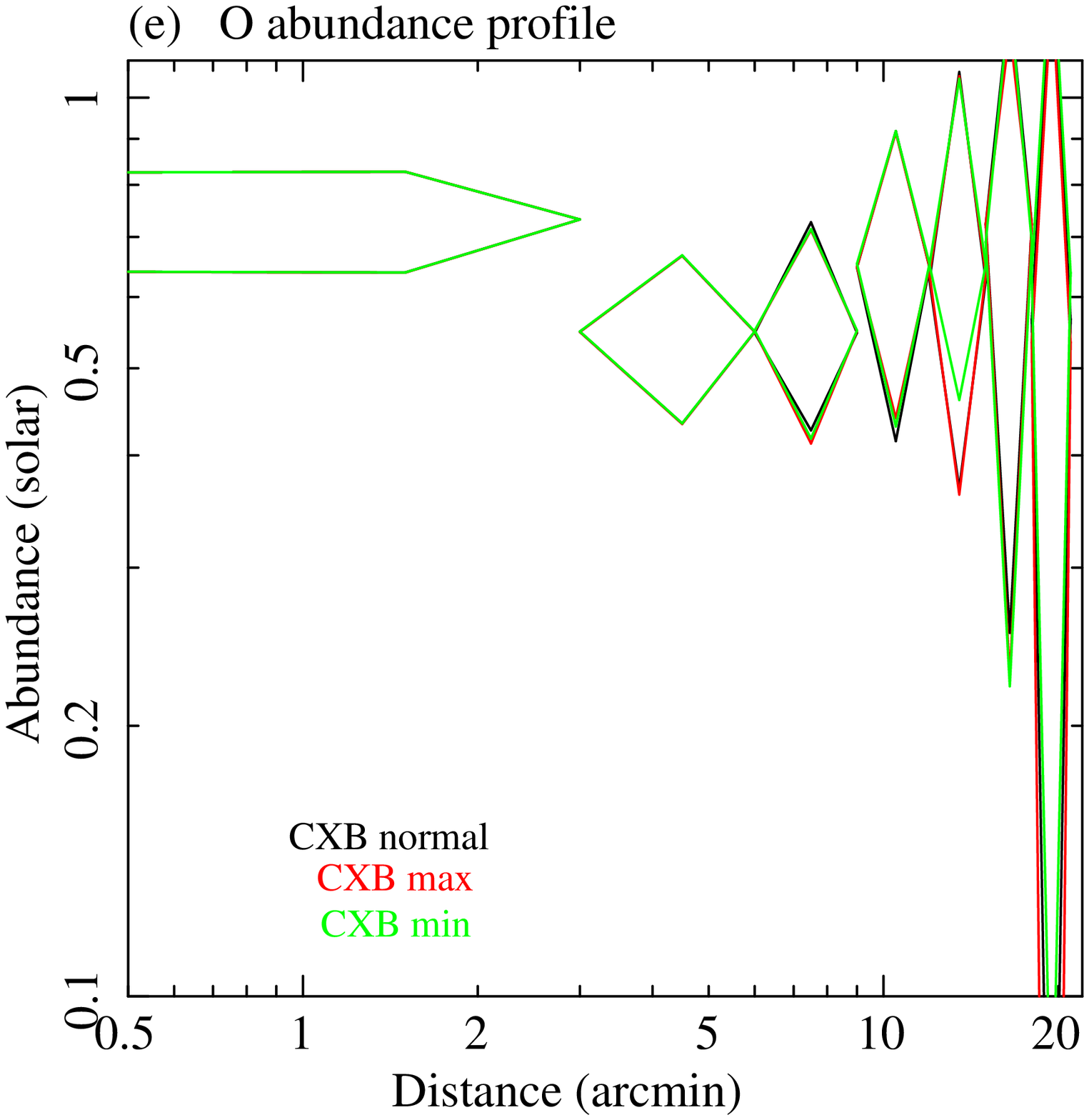} 
\hspace{0.3cm}
   \includegraphics[width=0.25\textwidth,angle=-90,clip]{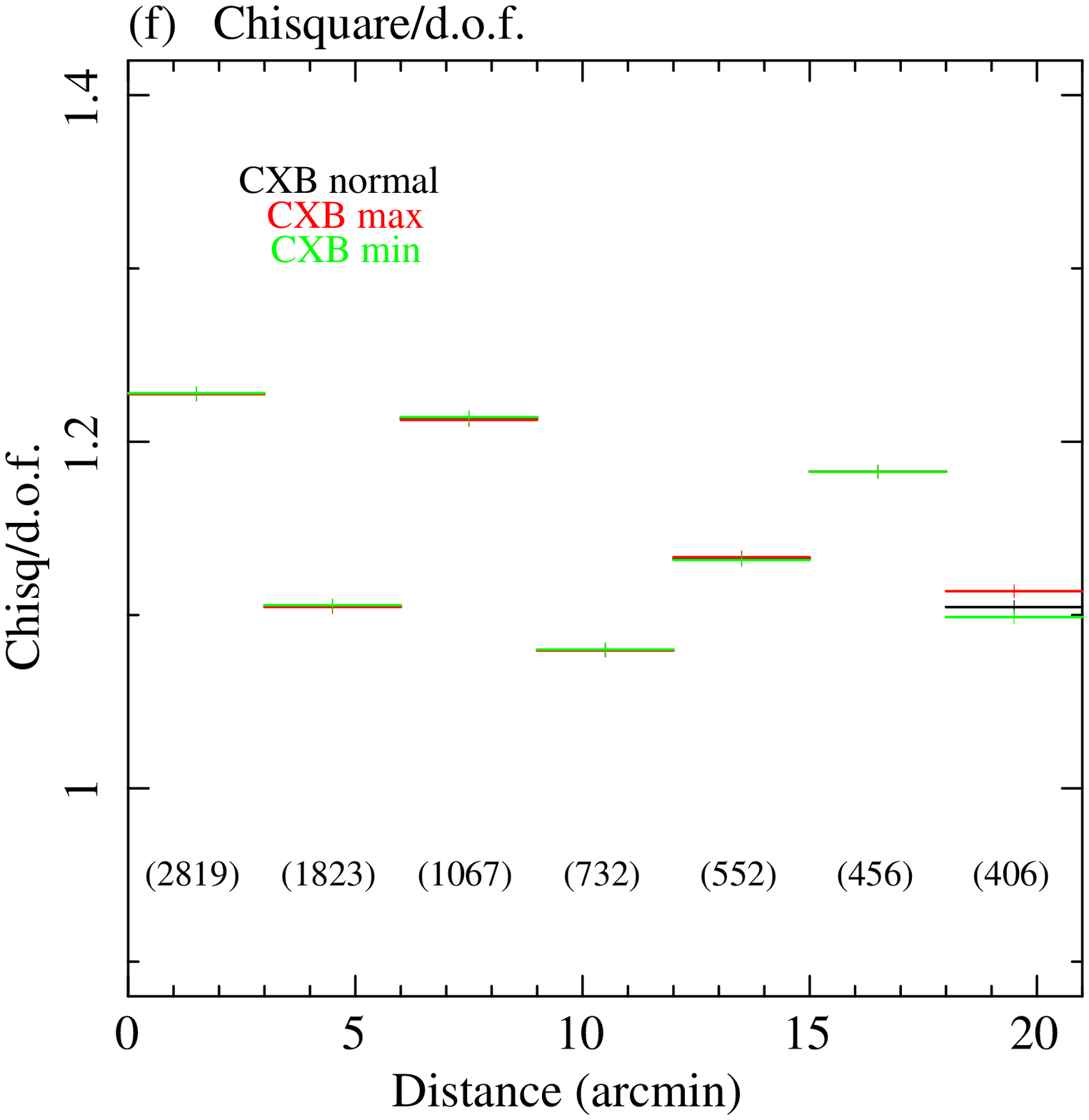} 
  \end{center}
  \caption{(a) The temperature profile of Abell~2199 determined by the single 
temperature {\it vapec} model (black). 
Those when the CXB level is set maximum and minimun are also shown in red and green, respectively.
(b) The same as panel (a), but for the hydrogen column density. The Galactic value is shown as 
a horizontal dotted line.
(c)-(e) Abundances of Fe, Si, and O, respectively, with the same color specifications as
panel (a). (f) Reduced chi-squared of the fit. 
The degree of freedom is also shown in the parentheses.
See the electronic version of the paper for colour figures. 
}
\label{fig:vapec-prof}
\end{figure}

A soft excess may manifest itself as a hydrogen column density $N_{\rm H}$  that is
significantly lower than the Galactic HI columnn density toward the Abell~2199 
field (weighted-averaged value over $1^{\circ}$ radius from NGC~6166),
$8.60 \times 10^{19}$ cm$^{-2}$ from \citet{dickey-1990}
or $8.92 \times 10^{19}$ cm$^{-2}$ from \citet{kalberla-2005}.
The measured values of $N_{\rm H}$ in 
figure~\ref{fig:vapec-prof} and figure~\ref{fig:2vapec-prof}, 
however, are factor of $2$--$3$ {\it higher} than the Galactic value in all regions. 
While this result generally argues against the presence of any flux excess in
lower energies, we need to understand the nature of the higher absorption before
discussing the soft excess issue.
There are five possibilities to explain this effect: (1) excess absorption 
within the Abell~2199 system, (2) an overestimate of the background which was
subtracted, (3) inadequate modeling of the ICM emission, (4) presence of 
high metallicity clouds at the direction of Abell~2199, and (5) underestimate of 
the XIS contaminant thickness which is included in the ARF. 
The first possibility is unlikely, because the measured $N_{\rm H}$ is high even 
in the outer regions such as
$12'$--$15'$. The second alternative is rejected, because we already incorporated 
the uncertainties of NXB and CXB in the analysis, and because within $6'$, the background 
level is more than an order of magnitude lower than the signal below 1 keV band.  
The third is also unrealistic, because the values of $N_{\rm H}$ remained unchanged
between the single- and two-temperature modelings. 
The forth is probably unphysical because metal abundances in such clouds would have to 
be as large as $\sim 3$ solar in order to explain the excess X-ray absorption, while
keeping  $N_{\rm H}$ to the Galactic HI value. Then, we focus on the last possibility.

\begin{figure}
  \begin{center}
   \includegraphics[width=0.25\textwidth,angle=-90,clip]{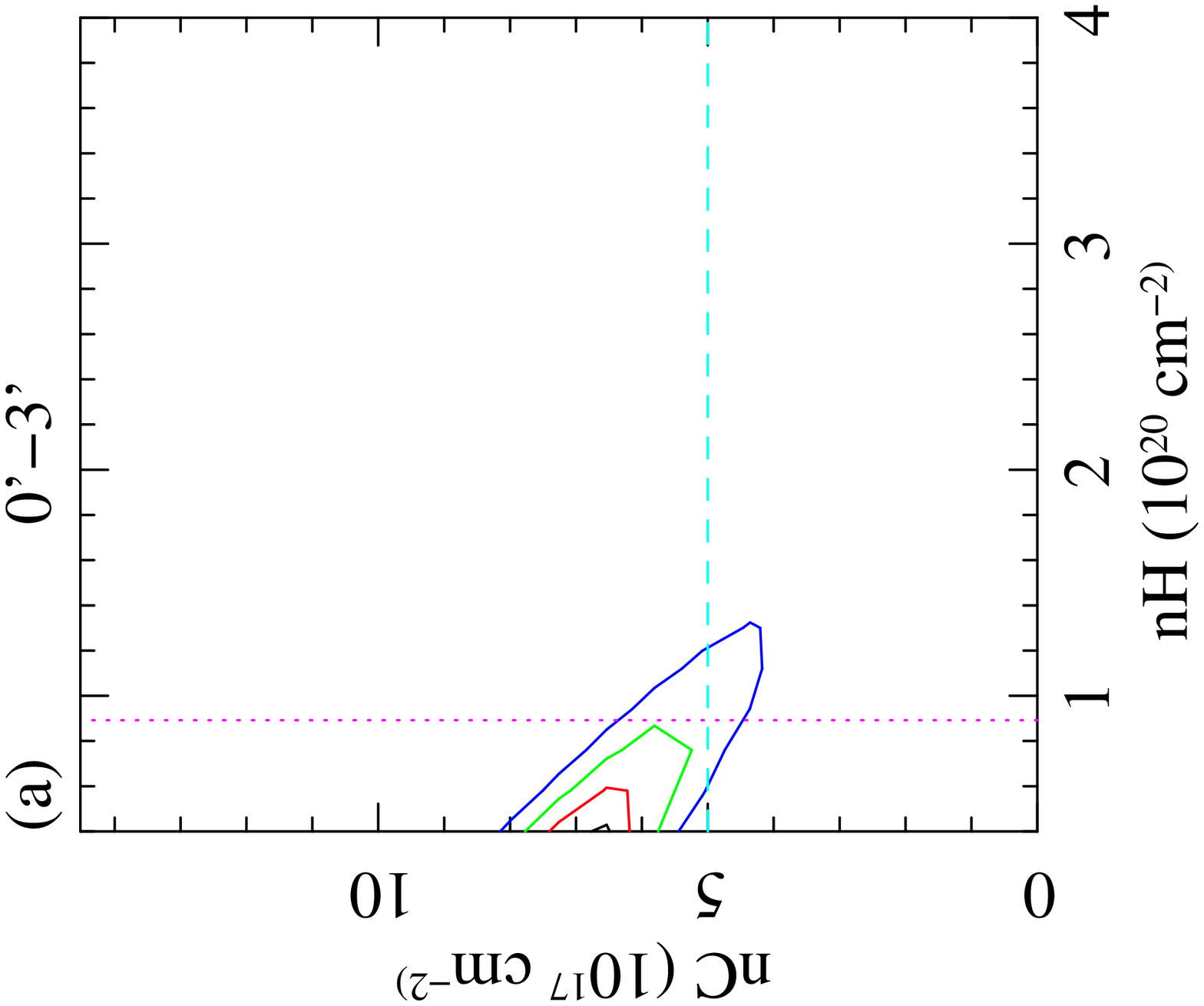} 
\hspace{0.2cm}
\vspace{0.2cm}
   \includegraphics[width=0.25\textwidth,angle=-90,clip]{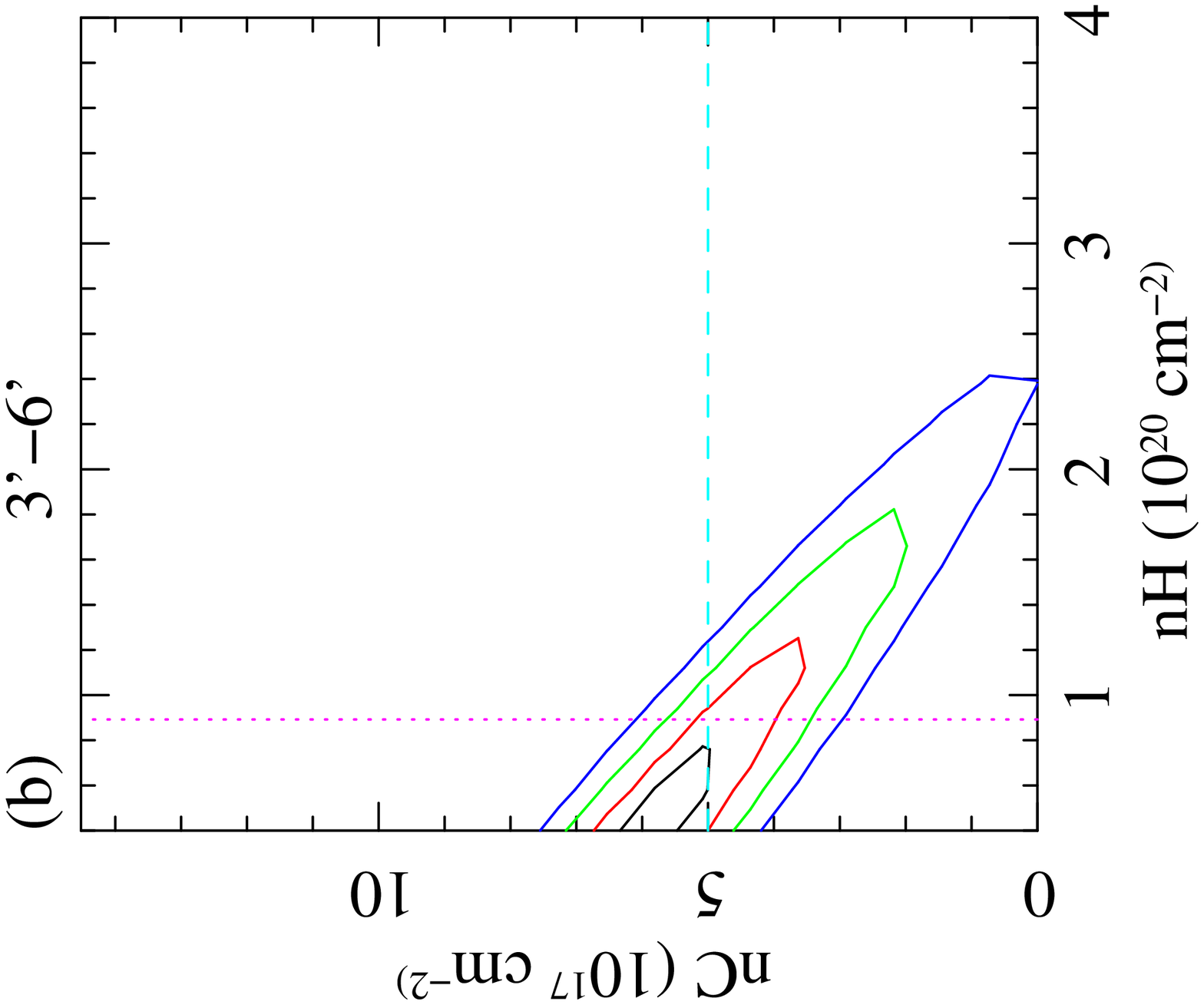} 
\hspace{0.2cm}
\vspace{0.2cm}
   \includegraphics[width=0.25\textwidth,angle=-90,clip]{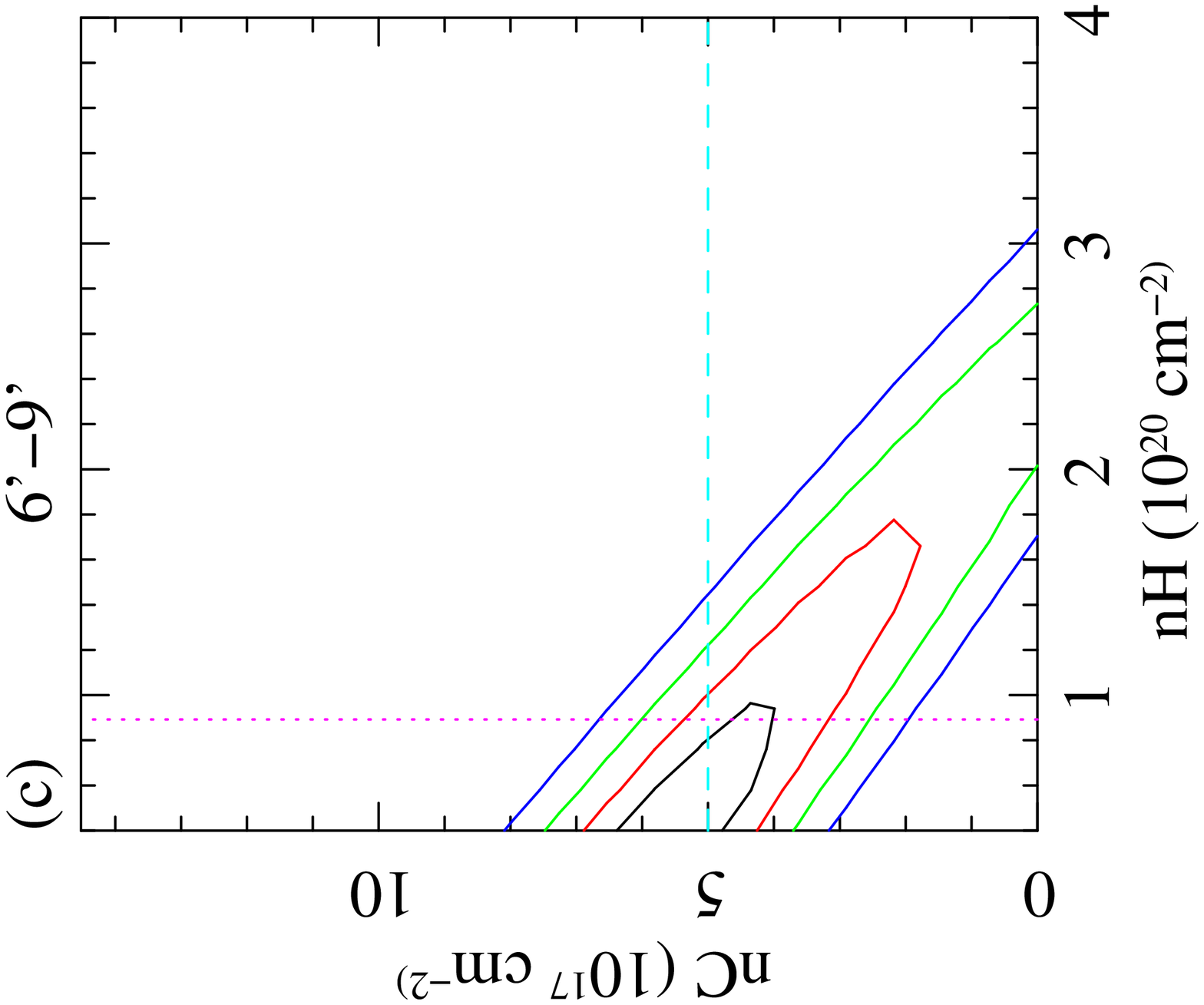} 
\hspace{0.2cm}
\vspace{0.2cm}
   \includegraphics[width=0.25\textwidth,angle=-90,clip]{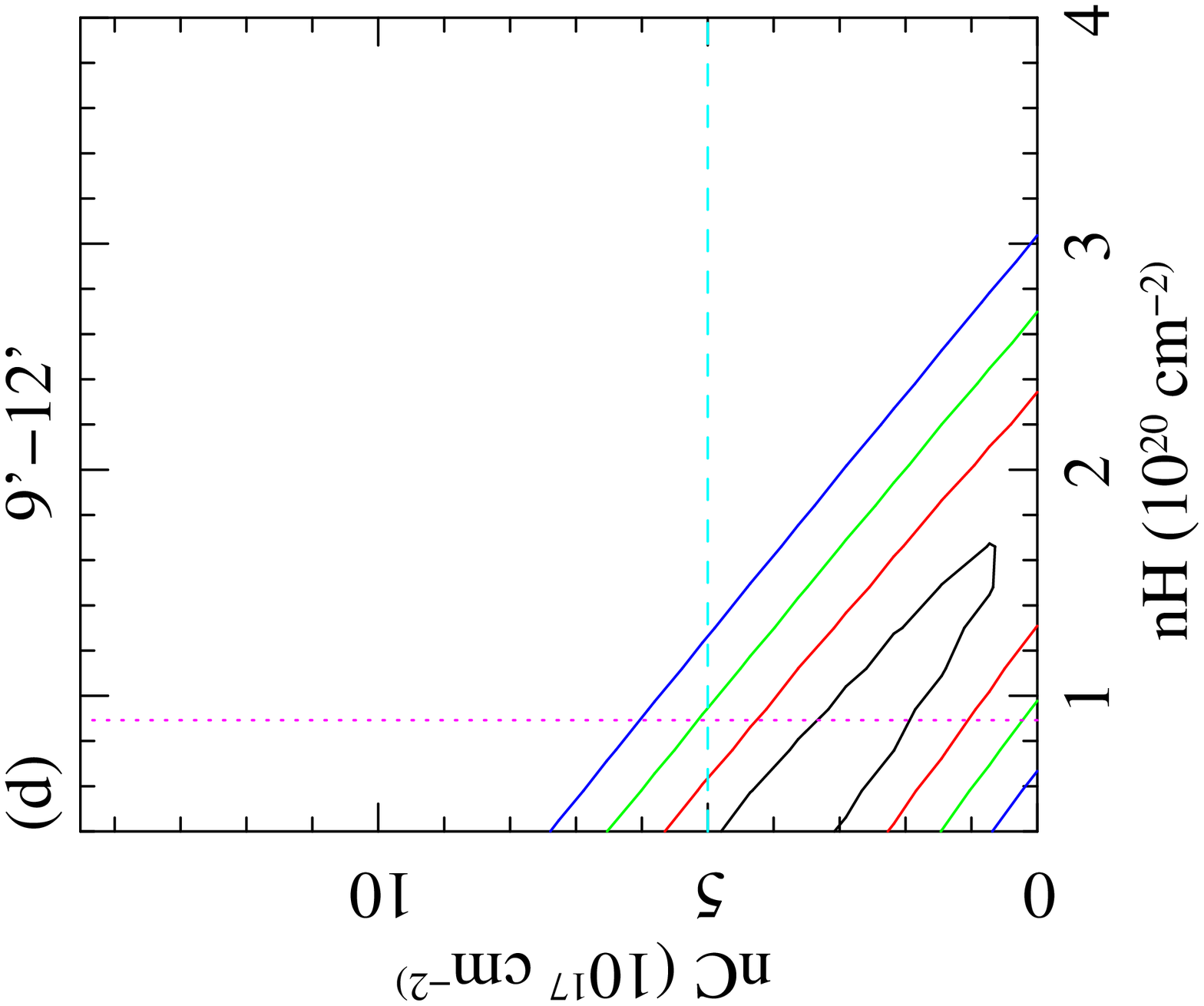} 
\hspace{0.2cm}
\vspace{0.2cm}
   \includegraphics[width=0.25\textwidth,angle=-90,clip]{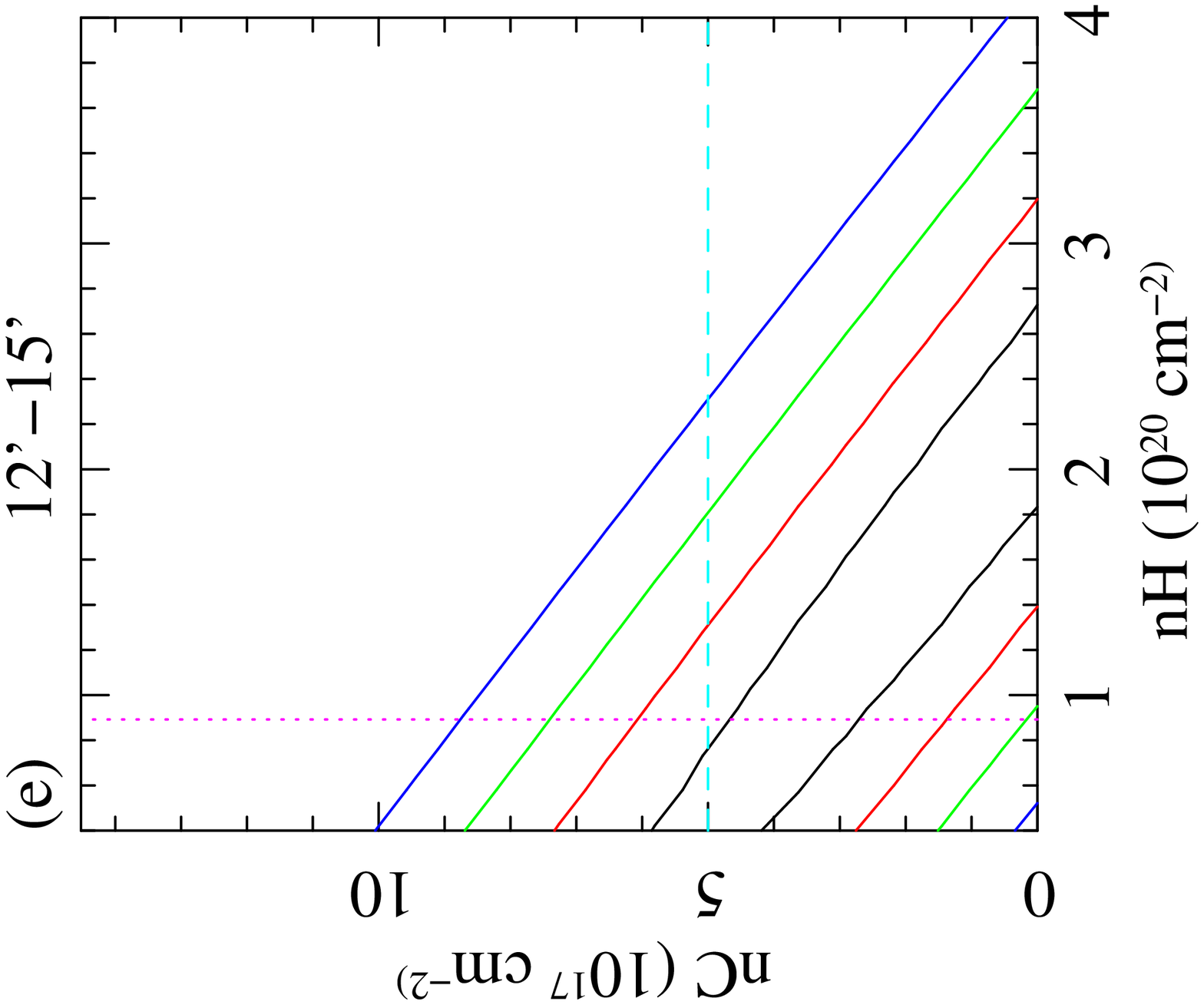} 
\hspace{0.2cm}
   \includegraphics[width=0.25\textwidth,angle=-90,clip]{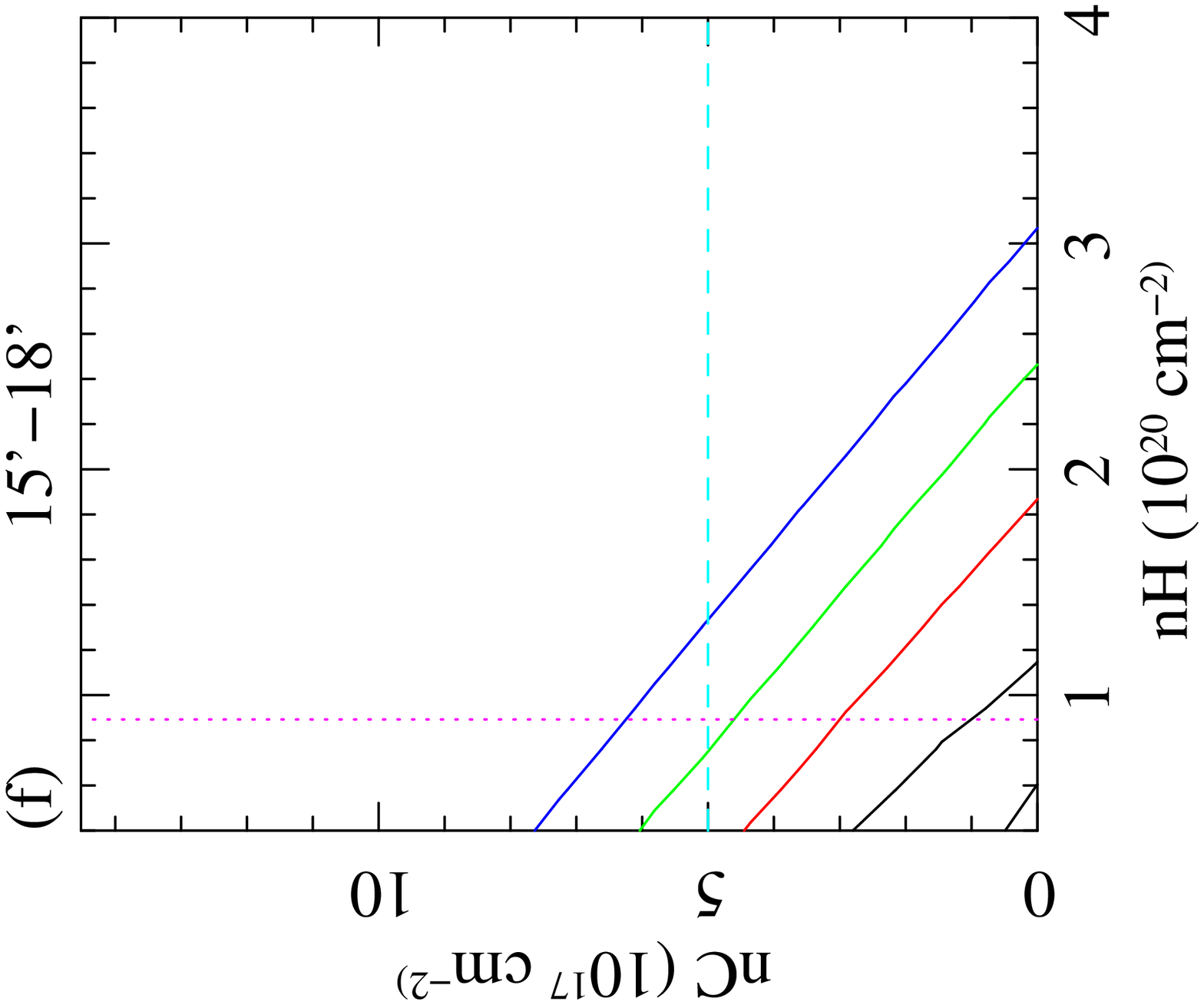} 
\hspace{0.2cm}
   \includegraphics[width=0.25\textwidth,angle=-90,clip]{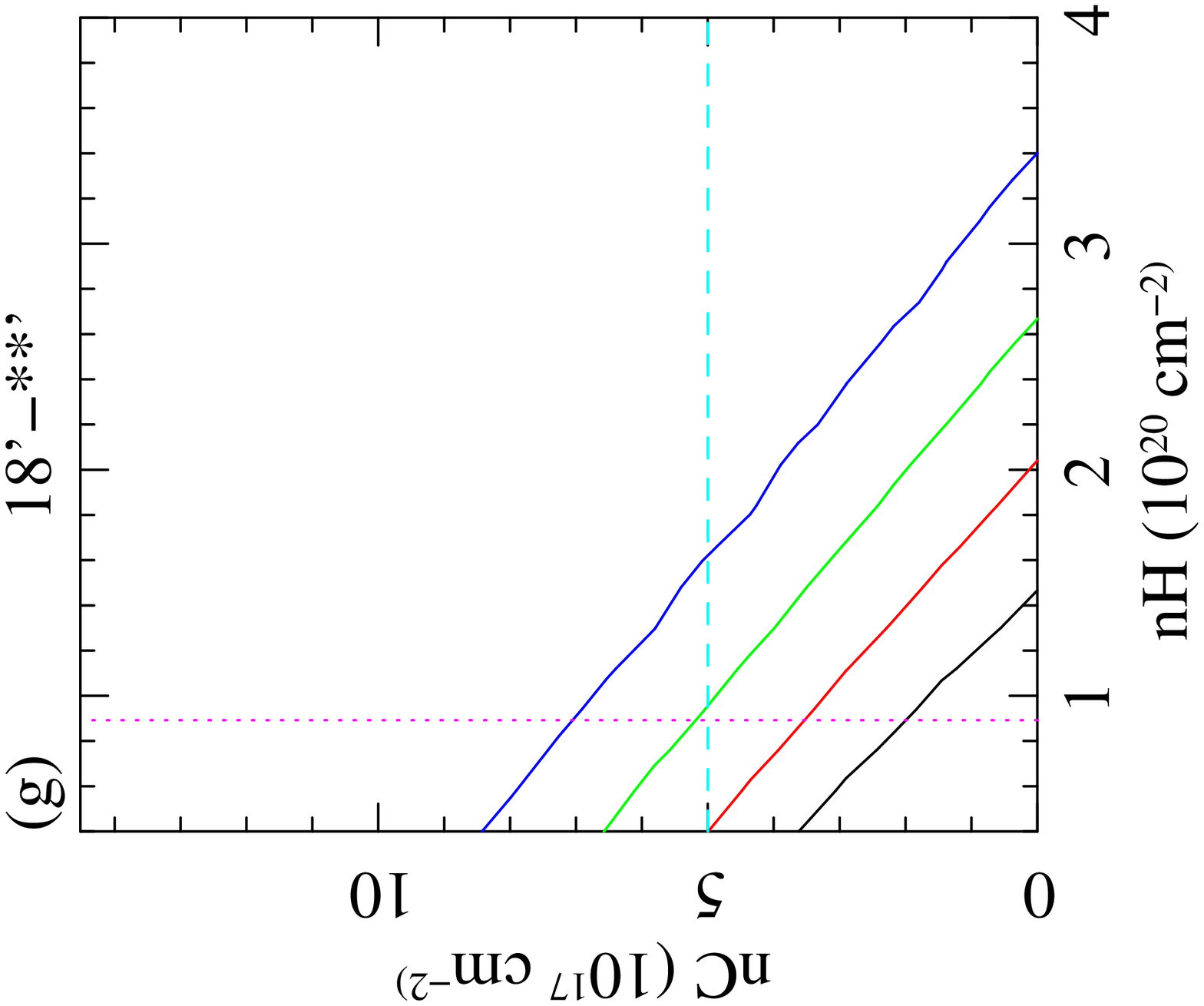} 
  \end{center}
  \caption{Confidence contours calculated for the seven regions
between the hydrogen column density $N_{\rm H}$, and the extra carbon contaminant
thickness $N_{\rm C}$ which is added to that already included in the nominal XIS ARFs.
Confidence levels are $1\sigma$ (black), $2\sigma$ (red), $3\sigma$ (green), 
and $4\sigma$ (blue). 
The Galactic HI columnn density of the Abell~2199 
field, $8.60 \times 10^{19}$ cm$^{-2}$ (\cite{dickey-1990}), is
shown in a dotted magenta line. A typical systematic uncertainty of the
contaminant thickness in the ARF, $5 \times 10^{17}$ cm$^{-2}$, is also shown
in a dashed cyan line. 
See the electronic version of the paper for colour figures. 
}
\label{fig:conf-contour}
\end{figure}

When we obtain $N_{\rm H}$ separately with XIS-FI and XIS-BI in the three inntermost
regions where the values of $N_{\rm H}$ are constrained well (figure~\ref{fig:vapec-prof}(b)),
the results with XIS-FI become systematically higher than those with XIS-BI by
$\sim 20$\%. This difference suggests an instrumental cause such as (5). 
Then, we evaluated an excess thickness of contaminant, 
assuming (5) is the cause of the high $N_{\rm H}$.
We included an additional absorption by the XIS contamination in the single 
temperature model as {\it vapec$\times$phabs$\times$varabs}. Here, the {\it varabs} factor
represents the additional absorption which is added to that already included in the ARF, 
and can handle separately column densities of elements from hydrogen to nickel.
The {\it varabs} column densities of all the elements except carbon and oxygen 
were fixed to zero, and the relative column-number-density ratio of carbon to oxygen
was fixed to $N_{\rm C}/N_{\rm O} = 6$ (\cite{koyama-2007}). 
By introducing the additional {\it varabs} factor, the best-fit model parameters did not
change significantly except $N_{\rm H}$ in {\it phabs}. 
Figure~\ref{fig:conf-contour} shows the confidence contours between
$N_{\rm H}$ in {\it phabs} and the additional carbon
column density $N_{\rm C}$  in {\it varabs}. As easily expected, these two quantities
correlate negatively with each other, and we cannot determine 
the true $N_{\rm H}$ accurately. Fixing $N_{\rm H}$ at the 
Galactic HI value in figure~\ref{fig:conf-contour} (dotted magenta lines) yields
$N_{\rm C} \sim 5 \times 10^{17}$ cm$^{-2}$ (dashed cyan lines in figure~\ref{fig:conf-contour}), 
which is comparable to an upper limit allowed by its systematic uncertainty\footnote{http://www.astro.isas.jaxa.jp/suzaku/process/caveats/caveats\_xrtxis08.html}. 
Therefore, we may regard the excess absorption observed in figure~\ref{fig:vapec-prof}b
and figure~\ref{fig:2vapec-prof}b as due to an underestimation of 
the XIS contaminant  by $N_{\rm C} \sim 5 \times 10^{17}$ cm$^{-2}$, and presume that
$N_{\rm H}$ is equal to the Galactic HI value. 
From figure~\ref{fig:conf-contour} of the innermost annuli,
we can reject any $N_{\rm H}$ value which is significantly lower than the Galactic HI value 
(dotted magenta lines in figure~\ref{fig:conf-contour}),
because the extra carbon thickness must be $< 5 \times 10^{17}$ cm$^{-2}$.

\begin{figure}
  \begin{center}
   \includegraphics[width=0.3\textwidth,angle=0,clip]{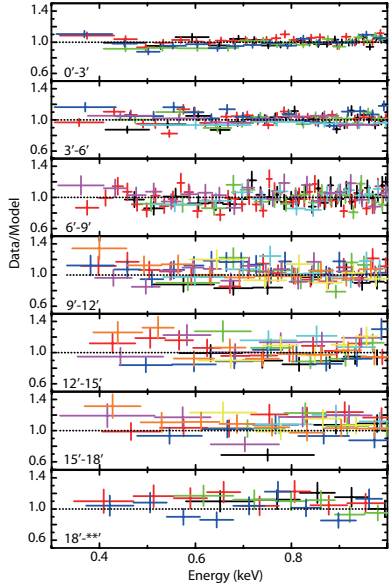} 
  \end{center}
  \caption{Data-to-model spectral ratios in the 0.3--1.0 keV range, 
when the data in each annulus are fitted with a single temperature \texttt{vapec} model 
with fixed
hydrogen ($N_{\rm H} = 8.92 \times 10^{19}$ cm$^{-2}$)
and carbon ($N_{\rm c} = 5.0 \times 10^{17}$ cm$^{-2}$)
column densities.
From top to bottom, the ratio spectra of the $0'$--$3'$, $3'$--$6'$, $6'$--$9'$, 
$9'$--$12'$, $12'$--$15'$, $15'$--$18'$, and $18'$--$21'$ regions are shown. 
Colors specify different observations. 
See the electronic version of the paper for a colour figure. 
}
\label{fig:resi-spec}
\end{figure}

In figure~\ref{fig:resi-spec}, we show data-to-model ratios in the 0.3--1.0 keV range,
when the data are fitted by the single temperature {\it vapec} model
with $N_{\rm H} = 8.60 \times 10^{19}$ cm$^{-2}$ 
and $N_{\rm C} = 5.0 \times 10^{17}$ cm$^{-2}$ both fixed.
The value of $N_{\rm C}$ is thus set to the allowed maximum, and the 
combination of $N_{\rm C}$ and $N_{\rm H}$ means a slight ($\sim 2.5 \sigma$)
overestimate of the overall absorption specified by the data
in figure~\ref{fig:conf-contour} of the innermost annulus. 
Furthermore, the data are divided by the single-temperature model which was
determined over a wide band of $0.3$--$8.0$ keV. Therefore, soft excess emission,
if any, should be seen in these ratio spectra.
However, no significant systematic excess is seen therein.
Then, to obtain an upper limit on the soft excess due to a thermal warm gas, 
we added to the model a 0.2 keV {\it apec} component with 0.2 solar metallicity,
following the result of \citet{kaastra-2004}. 
The derived 90\% upper limit on surface brightness of the 0.2 keV component is 
summarized in table~\ref{table:warm-upper}.
In table~\ref{table:warm-upper}, we also give XIS upper limits
when a power law with photon index of 2.0 is added to the model.
The upper limit on the power law in terms of the 0.2--10 keV luminosity becomes
$7.1 \times 10^{43}$ erg s$^{-1}$ ($2.5 \times 10^{43}$ erg s$^{-1}$ in 20--80 keV range)
when integrated to $30'$. In subsection~\ref{subsection:results:hard-excess},  
we independently constrain such a power law component using the HXD-PIN data.

As a cross confirmation, we repeated the same analysis by choosing 
$N_{\rm C} = 0 $ and $N_{\rm H} = 3 \times 10^{20}$ cm$^{-2}$, the latter
suggested by figure~\ref{fig:vapec-prof}. However, the ratio spectra remained
essentially unchanged. 






\subsection{\it Constraints on hard excess emission
\label{subsection:results:hard-excess}}
\begin{figure}
  \begin{center}
   \includegraphics[width=0.3\textwidth,angle=0,clip]{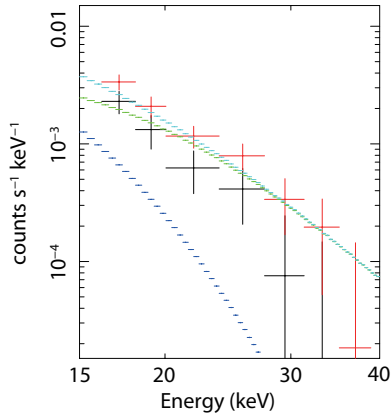} 
  \end{center}
  \caption{NXB-subtracted 15--40 keV PIN spectra (black) summed over
the Center, Offset1, Offset2, and Offset4 observations. 
The same spectrum when the NXB is lowered within the 90\% confidence limit
is also plotted in red. The CXB model (green), the extrapolated thermal emission (blue), 
and their sum (cyan) are also plotted for comparison.
}
\label{fig:pin-spec}
\end{figure}

In figure~\ref{fig:pin-spec}, we show a 15--40 keV NXB-subtracted HXD-PIN spectrum 
summed over Center, Offset1, Offset2, and Offset4. By this summation, 
the statistical error associated with the 15--40 keV HXD-PIN signal reduced to $\sim 0.8$\%
of the NXB, which is considerably smaller than the systematic NXB error of $\pm 2.3$\%
(90\% confidence range) adopted in the present paper. 

After subtracting the NXB, the HXD-PIN signal (black crosses in figure~\ref{fig:pin-spec}) 
was thus detected significantly up to $\sim 30$ keV. For comparison, we also show in 
figure~\ref{fig:pin-spec} spectra of the CXB and the thermal emission.
Here, the thermal emission was modeled by summing the single 
temprature {\it vapec} models determined for individual annular regions.
As to thermal emission outside  the XIS FOV which contributes $\sim 11$\%, 
we assumed that the surface brightness obeys the double-$\beta$ 
profile which was described in \S~\ref{subsection:obs:xis},
and that the temperature and metal 
abundances therein is the same as those in the outermost 
region covered by the XIS. 
After the nominal XIS vs. HXD-PIN cross normalization (\cite{kokubun-2007}, 
see also a Suzaku calibration document named suzakumemo-2007-11\footnote{http://www.astro.isas.jaxa.jp/suzaku/doc/suzakumemo/suzakumemo-2007-11.pdf}), 
the model normalization of the thermal component was multiplied by a factor of 1.13
when calculating its predicted contribution to the HXD-PIN data. 
Using this cross normalization, 
intensities of the two CXB models in 6.0--12.0 keV agree within
1\% as described in \S~\ref{subsection:bgd:pin}.
This assures that the systematic errors associated with the cross normalization
is $< 1$\%, which is much smaller than the statistic erros in the HXD-PIN spectrum.
Thus, the NXB-subtracted spectrum is consistent, within statistical errors,
with a sum of the CXB and the extrapolated thermal emission (cyan points in figure~\ref{fig:pin-spec}). 
Even when the NXB is set to the minimum that is allowed by the 90\% systematic 
uncertainty range, the HXD-PIN signal does not exceed the (CXB+thermal)
emission significantly. Thus, the hard X-ray excess was not detected.

In order to obtain an upper limit on hard excess emission,
we lowered the NXB within its 90\% systematic uncertainty, and
also lowered the CXB model within its fluctuation of 18\% (90\%-confidence).  
The estimation of the CXB fluctuation is described in 
Appendix~\ref{appen:cxb-sys}. Then, we fitted the 
NXB-subtracted spectrum with a power law (with its photon index fixed
at 2.0 and its normalization left free) plus the fixed CXB and 
thermal models, and obtained upper limits. 
This very conservative method gave a 90\%-confidence upper limit of 
$1.8 \times 10^{43}$ erg s$^{-1}$ on the 20--80 keV luminosity of the
power-law component. If this component is uniformly distributed over a circular
region of radius $10'$, the implied 0.2--10 keV luminosity per solid angle
becomes $1.6 \times 10^{41}$ erg s$^{-1}$ arcmin$^{-2}$. This limit is comparable
to those obtained with the XIS using the softest specctral end.


\section{Discussion
\label{section:discussion}}

Using the five pointing observations with Suzaku, we measured the temperature and
metal abundances of Abell~2199 out to $\sim 700 $ kpc (0.4 times virial radius), and
searched the XIS and HXD-PIN data for soft and hard excess emissions, respectively, 
both above the thermal component.
We however found no significant excess emissions in either energy bands, 
and derived their upper limits. Over a broad energy band of $\sim 0.4$ keV to 
$\sim 30$ keV, the emission is dominated by the thermal components with temperatures
of a few keV. 

The upper limits on soft excess, expressed as the emission measure per unit solid
angle from a 0.2 keV warm gas 
(table~\ref{table:warm-upper}), are more stringent, typically by more than a factor of 3, 
than the reported positive detection by XMM-Newton (\cite{kaastra-2004}),
$(28 \pm 13) \times 10^{62}$ cm$^{-3}$ arcmin$^{-2}$. 
One likely cause of this inconsistency between Suzaku and XMM-Newton is
confusion with solar wind charge exchange (SWCX) emission, which 
has been observed with Chandra (\cite{wargelin-2004}), XMM-Newton (\cite{snowden-2004}), 
and Suzaku (\cite{fujimoto-2007}).
In fact, proton flux near the Earth, observed with 
WIND-SWE\footnote{http://web.mit.edu/space/www/wind.html}, enhanced by a factor 
of 3--4 for a part of the XMM-Newton observatons compared with that of Suzaku. 
The relevant proton light curves are shown in Appendix~\ref{appen:proton-flux}.
Similarly, in Suzaku observations of outskirt regions of the Coma cluster, 
Takei et al. (2008) found no significant excess
O~VII or O~VIII emissions which were previously reported with XMM-Newton (\cite{finoguenov-2003}), 
and concluded that the excess emission is likely due to an enhancement in the SWCX 
emission during the XMM-Newton observations. 


The Suzaku data have constrained the 20--80 keV luminosity of any excess
hard emission to be $< 1.8 \times 10^{43}$ erg s$^{-1}$, assuming that
it has a power-law spectrum of photon index 2.0 and employing the most
conservative estimates of the CXB.
Using the BeppoSAX PDS, \citet{nevalainen-2004} detected non-thermal 
emission  from Abell~2199 at 2.1$\sigma$ level, with 20--80 keV luminosity
of $(1.7 \pm 1.3) \times 10^{43}$ erg s$^{-1}$ (90\%-confidence errors)
assuming a power-law spectrum of photon index 2.0. 
This detection by the PDS is not rejected by our result.
However, the BeppoSAX results, with a significance of only 2.1$\sigma$,
would be considered as an upper limit, like the present result. 
In this case, our upper limit is a factor of 1.7 more stringent than that of the PDS. 
From a calibration analysis using the Crab Nebula, the 20--80 keV flux derived from 
the PDS is known to be systematically lower by 21\% than that derived from HXD-PIN 
(\cite{nakazawa-2009}). If we take into account this systematic effect, the difference 
in the upper limits between HXD-PIN and PDS increases to a factor of 2.1. 

Considering the difference between the FOVs of HXD-PIN ($34'$ in FWHM) 
and PDS ($1.3^{\circ}$ in FWHM), the claimed detection by the PDS becomes consistent
with our upper limit if the excess hard X-ray emission is much more extended than
the HXD-PIN FOV.
However, the total FOV of HXD-PIN in our pointing observations ($\sim 43'= 1.5$ Mpc 
radius in FWHM) mostly cover the whole cluster region, $\sim 0.9$ times the Virial 
radius of Abell~2199  ($\sim 1.7$ Mpc, see \S~\ref{subsection:bgd:sys}).
Therefore, to reconcile the two results, we would have to assume that
the emission is spatially distributed much beyond the virial radius.

\begin{figure}
  \begin{center}
   \includegraphics[width=0.38\textwidth,angle=0,clip]{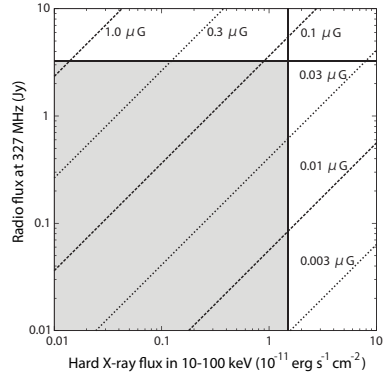} 
  \end{center}
  \caption{
Relations between the 10--100 keV hard X-ray flux and the 327 MHz flux density of 
Abell~2199, produced when the same population of relativistic electrons
with index 2.6 interact with the cosmic microwave background photons and
magnetic fields of various strengths, respectively. 
The upper limits on the radio (\cite{kempner-2000}) and hard X-ray (this work)
signals are shown in solid lines, and the gray region satisfies these limits. 
}
\label{fig:mag-limit}
\end{figure}

According to Kempner \& Sarazin (2000), a radio synchrotron flux density at 327 MHz, $S_{\nu}$, 
expected from Abell~2199, is related to 
the 10--100 keV IC X-ray flux, $S_{\rm HXR}$, as
\begin{equation}
S_{\nu} = 234 \;\left( \frac{S_{\rm HXR}}{10^{-11}\; {\rm erg}\; {\rm cm}^{-2}\; {\rm s}^{-1}} \right) \left( \frac{B}{1\;\mu {\rm G}} \right)^{1.81} \left( \frac{\nu}{327\; {\rm MHz}} \right)^{-0.81} {\rm Jy},
\label{eq:}
\end{equation}
assuming the photon index of the hard X-ray emission to be 1.81. Here, 
$B$ is the cluster magnetic field, and $\nu$ is the observed frequency. 
The upper limit they quoted on the diffuse radio flux, $S_{\nu} < 3.25$ Jy at 327 MHz, and 
the hard X-ray emission detected by BeppoSAX, required a very weak magnetic 
field ($< 0.073\; \mu$G). 
However, given our non-detection of such an excess hard X-ray flux, this 
is no longer the case.
When the HXD-PIN data are used to constrain a power-law component with a
photon index of 1.81 (instead of 2.0), its 10--100 keV luminosity becomes
$< 1.5 \times 10^{-11}$ erg cm$^{-2}$ s$^{-1}$.
As a result, effectively any strength of magnetic field  is allowed
as shown in figure~\ref{fig:mag-limit}.

Regardless of the magnetic field strength, the present result constrain
the number of relativistic electrons in the system, so that their IC emission
should not exceed the present hard X-ray upper limit.
If we assume a power-law spectrum of synchrotron emission with a photon index 2.0, 
the corresponding electron number spectrum has a power-law index 3.0, and is
written as \citep{petrosian-2006}
\begin{equation}
N(\gamma) = 2 N_{\rm tot} \gamma^{2}_{\rm min} \gamma^{-3} \; \; \;  (\gamma > \gamma_{\rm min}), 
\label{eq:}
\end{equation}
where $\gamma$ is the Lorentz factor, 
$N_{\rm tot}$ is the total number of electrons, and $\gamma_{\rm min}$ is a lower-cutoff in $\gamma$
which is conservatively estimated to be $\sim 10^3$ (\cite{petrosian-2006}). These relativistic
electrons produce IC emission with a 20--80 keV luminosity of
\begin{equation}
L_{\rm HXR} = 1.0 \times 10^{45} \left( \frac{N_{\rm tot}}{10^{65}}\right) (1+z)^4 \; {\rm erg}\;  {\rm s}^{-1}, \label{eqn:hxr-lumi}
\end{equation}
where $z$ is the redshift. From equation~\ref{eqn:hxr-lumi} and our upper limit, we obtain
$N_{\rm tot} <  1.6 \times 10^{63}$.  
Then, the integrated electron kinetic energy $K_{\rm e}$ for $\gamma > \gamma_{\rm min}$ in Abell~2199
becomes
\begin{equation}
K_{\rm e} = N_{\rm tot} m_{\rm e} c^2 (2 \gamma_{\rm min} - 1) < 2.6 \times 10^{60} \left( \frac{\gamma_{\rm min}}{10^{3}}\right) \; {\rm erg}, 
\end{equation}
where $m_{\rm e}$ is the electron mass. Since electrons 
with $10^3 \lesssim  \gamma \lesssim 10^4$, which contribute to the $20$--$80$ keV IC emission, 
have a liftime of $\sim 10^{8-9} $ yr in typical ICM conditions (\cite{petrosian-2001}), 
the acceleration luminosity in Abell~2199 can be constrained as 
$< 8.4 \times 10^{44} \; (\gamma_{\rm min}/10^3)  $ erg s$^{-1}$.
This upper limit is rather loose, and is three times higher than 
the 0.5--10 keV luminosity of the thermal emission.
If we assume that the magnetic field of Abell~2199 is $\sim 1\; \mu$G 
as suggested by Faraday rotation measurements (\cite{ge-1994}), 
the radio upper limit gives the dominant constraint on the relativistic
electrons, with the predicted  $S_{\rm HXR}$ two orders of magnitude below
the present upper limit (figure~\ref{fig:mag-limit}). 
In this case, the constraint of acceleration luminosity becomes 
$\lesssim 10^{43} \; (\gamma_{\rm min}/10^3)  $ erg s$^{-1}$ which is 
about an order of magnitude lower than the 0.5--10 keV luminosity.

The total mass of Abell~2199 is $1.3 \times 10^{14} M_{\odot}$ when integrated to
0.8 Mpc (\cite{fukazawa-2004}).
If Abell~2199 was formed via a major merger between two clusters with similar masses $M$
($\sim 7 \times 10^{13} M_{\odot}$ each) at a relative speed of $v \sim 3000$ km s$^{-1}$,
the kinetic energy deposited onto the merged system becomes
\begin{equation}
E_{\rm merger} \sim \frac{1}{2}Mv^2 =  6 \times 10^{63} \left( \frac{ M }{ 7 \times 10^{13} \; M_{\odot}} \right) \left( \frac{ v }{ 3000\; {\rm km}\; {\rm s}^{-1}} \right)^2\; {\rm erg}.
\end{equation}
Since shocks produced in cluster mergers are thought to last
typically $\tau \sim 10^9$ yr (\cite{takizawa-2000}), the energy input rate
to non-thermal electrons becomes
$\dot{E}_{\rm in} \sim 1 \times 10^{46} \; (f/0.05)$ erg s$^{-1}$, 
where $f$ is the fraction of energies given to them. 
Values of $\dot{E}_{\rm in} = 10^{46-47}$ erg s$^{-1}$ are also expected
from theoretical models of merging clusters such as Coma (e.g. \cite{takizawa-2004}).
Then, our upper limit on the acceleration luminosity is an order
of magnitude lower than  $\dot{E}_{\rm in}$,  assuming $f \sim 0.05$.
We hence infer that Abell~2199 have been free from major merger events
for more than $\sim 10^9$ yr, which is a typical time scale on which
the IC emission decays (\cite{takizawa-2004}).

%

MK acknowledges support from a grant based on the Special Postdoctoral 
Researchers Program of RIKEN. The present work is supported in part
by the Grant-in-Aid for Scientific Research (S), No. 18104004.

\appendix

\section{Estimation of NXB systematic error
\label{appen:nxb-sys}}
$1\sigma$ systematic erros of PIN 15--40 keV band, using earth occultation data 
and E0102-72 observations which has been regularly scheduled for the XIS calibration,
is 2.3\% (\cite{fukazawa-2009}). Since the statistical error
is 1.8\% ($1\sigma$), the systematic uncertainty of the PIN NXB model 
in 15--40 keV band is estimated to $\pm 2.3$\% in 90\% confidence region.

\section{Estimation of CXB systematic error
\label{appen:cxb-sys}}
The CXB has uncertainty because the number of
unresolved faint sources in FOV statistically fluctuate.  
This CXB fluctuation scales as $\Omega_{\rm e}^{-0.5} S_{\rm c}^{0.25}$,
where $\Omega_{\rm e}$ is the FOV and $S_{\rm c}$ is the
detection threshold flux for point sources (\cite{ishisaki-1997}). 
By scaling from the CXB fluctuation of HEAO-I, 2.8\% for 
$\Omega_{\rm e} = 15.8$ deg$^{2}$ and 
$S_{\rm c} = 8 \times 10^{-11}$ erg s$^{-1}$ cm$^{-2}$ 
(\cite{shafer-1983}), that of the HXD is estimated
to be 18\%, assuming $\Omega_{\rm e} = 0.32$ deg$^{2}$ and
$S_{\rm c} = 8 \times 10^{-12}$ erg s$^{-1}$ cm$^{-2}$ in the
10--40 keV band. This corresponds to 1.3\% of the NXB level. 

\section{Normalization of the background model
\label{appen:bgd-norm}}
The background model (GFE+CXB) are determined separately for the seven annular regions. 
We show the normalizations of background components in figure~\ref{fig:bgd-norm}.
The normalizations of background models are consistent among the regions within errors
(90\% confidence range).

\section{Proton flux of Abell~2199 observations
\label{appen:proton-flux}}
\begin{figure}
  \begin{center}
   \includegraphics[width=0.2\textwidth,angle=-90,clip]{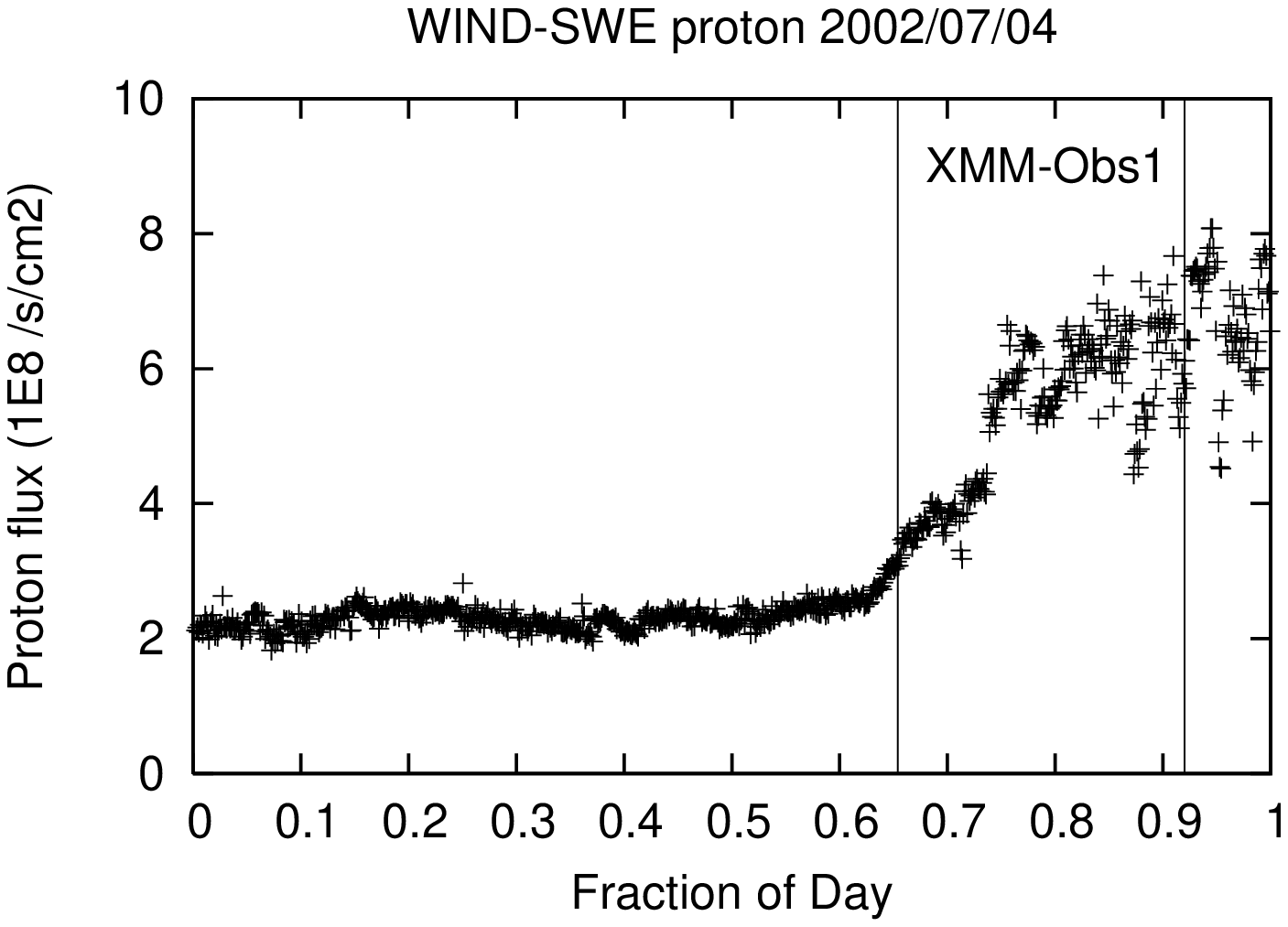} 
\hspace{0.2cm}
\vspace{0.2cm}
   \includegraphics[width=0.2\textwidth,angle=-90,clip]{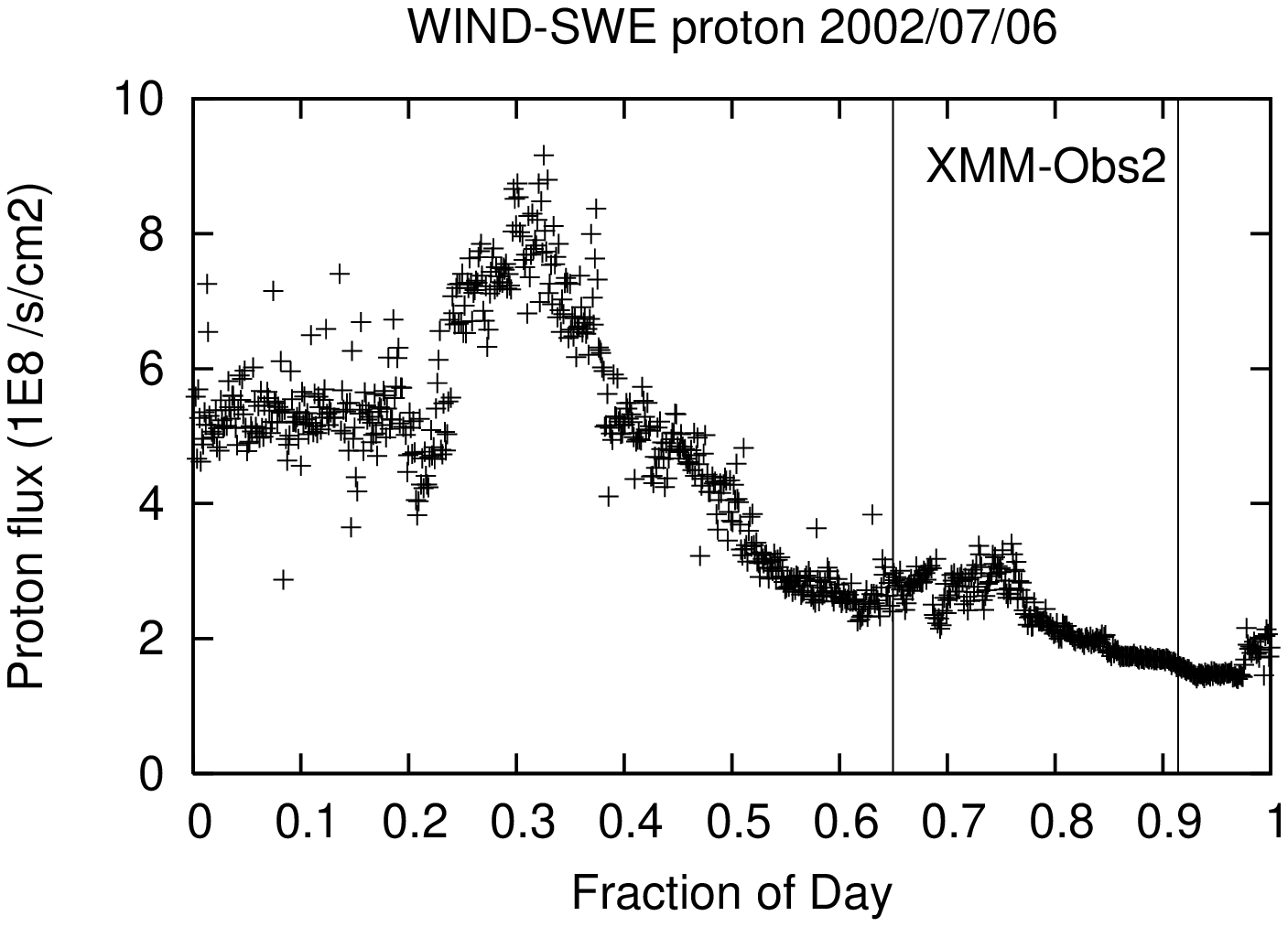} 
\hspace{0.2cm}
\vspace{0.2cm}
   \includegraphics[width=0.2\textwidth,angle=-90,clip]{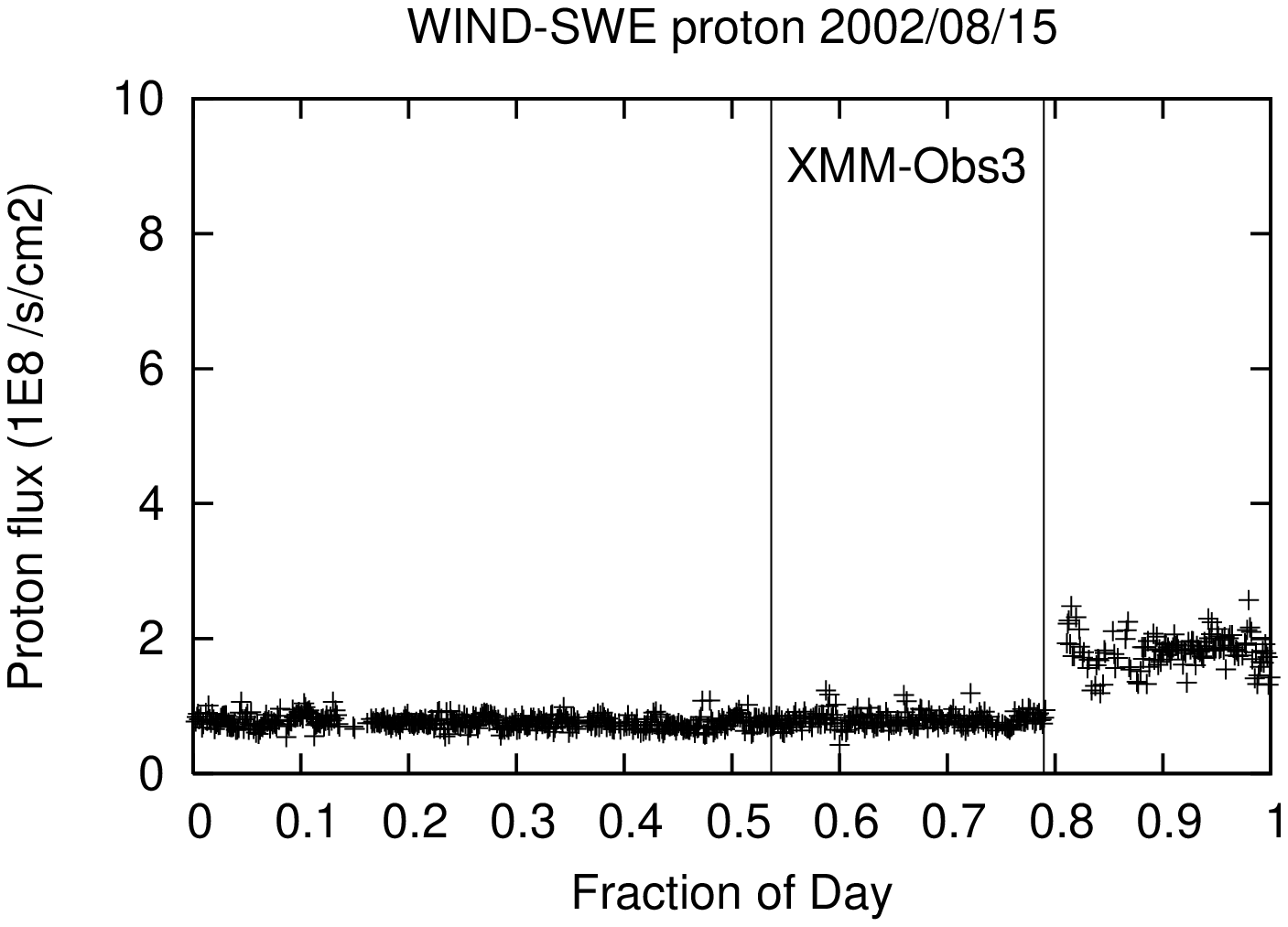} 
\hspace{0.2cm}
\vspace{0.2cm}
   \includegraphics[width=0.2\textwidth,angle=-90,clip]{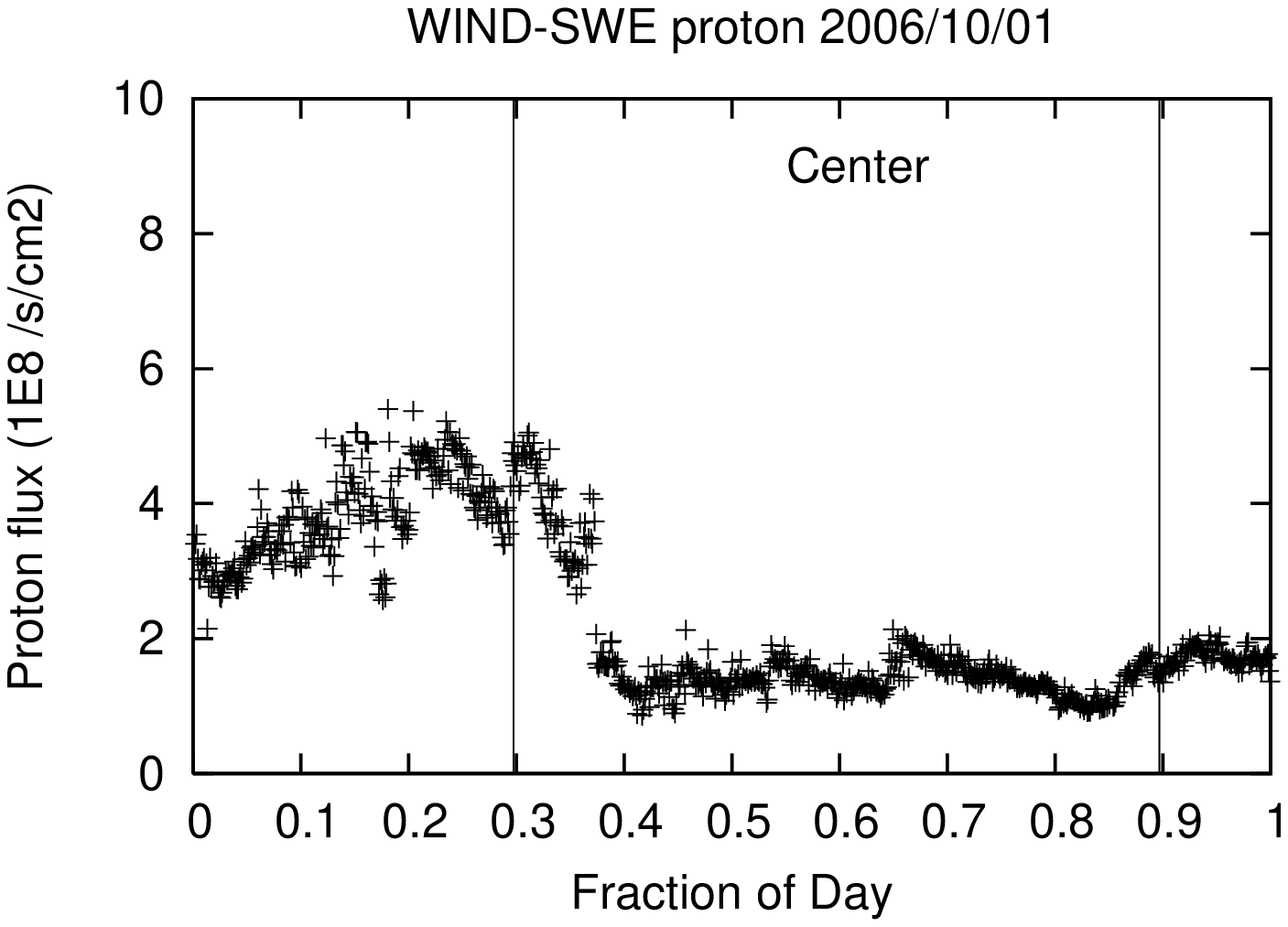} 
\hspace{0.2cm}
\vspace{0.2cm}
   \includegraphics[width=0.2\textwidth,angle=-90,clip]{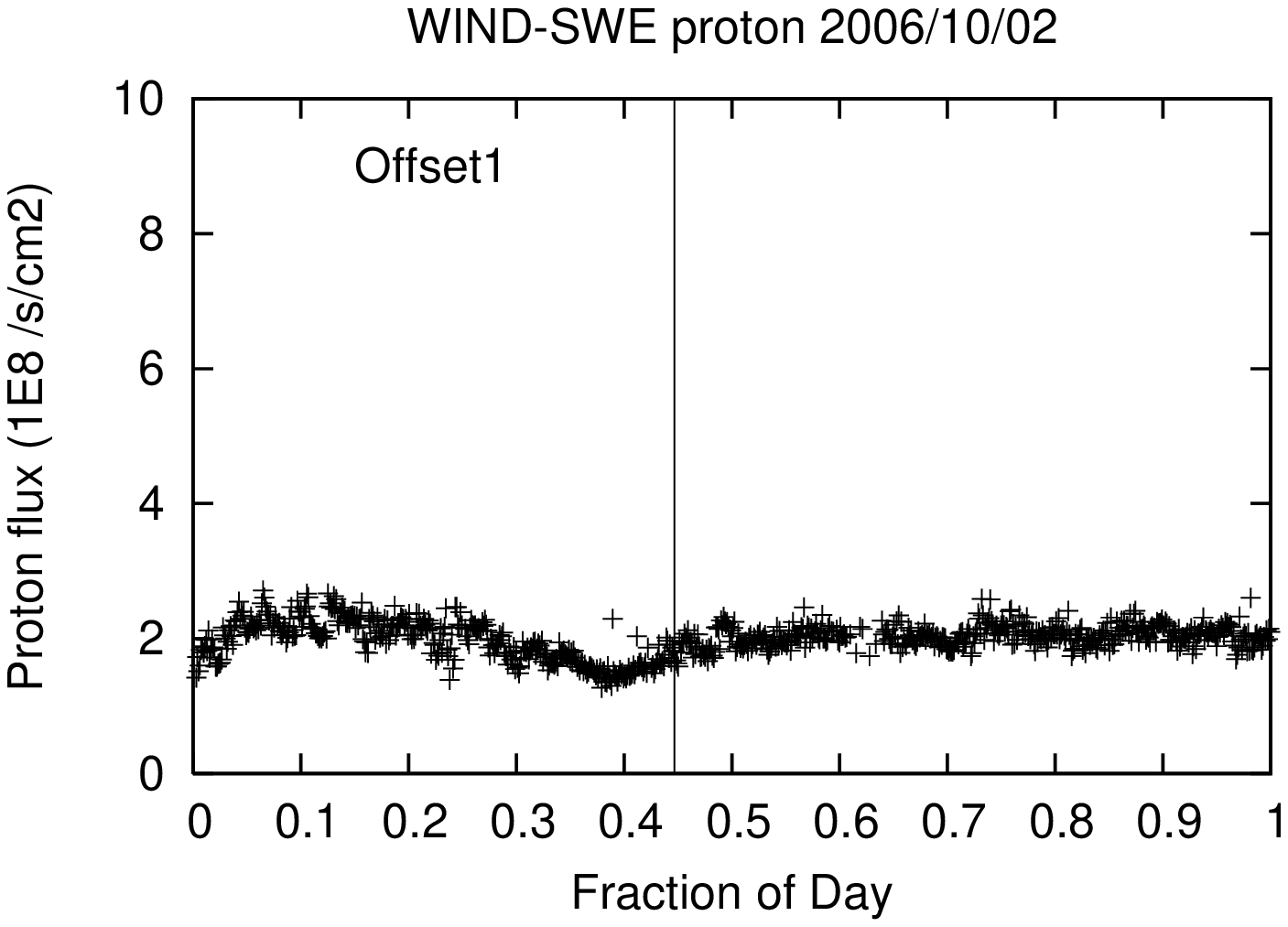} 
\hspace{0.2cm}
\vspace{0.2cm}
   \includegraphics[width=0.2\textwidth,angle=-90,clip]{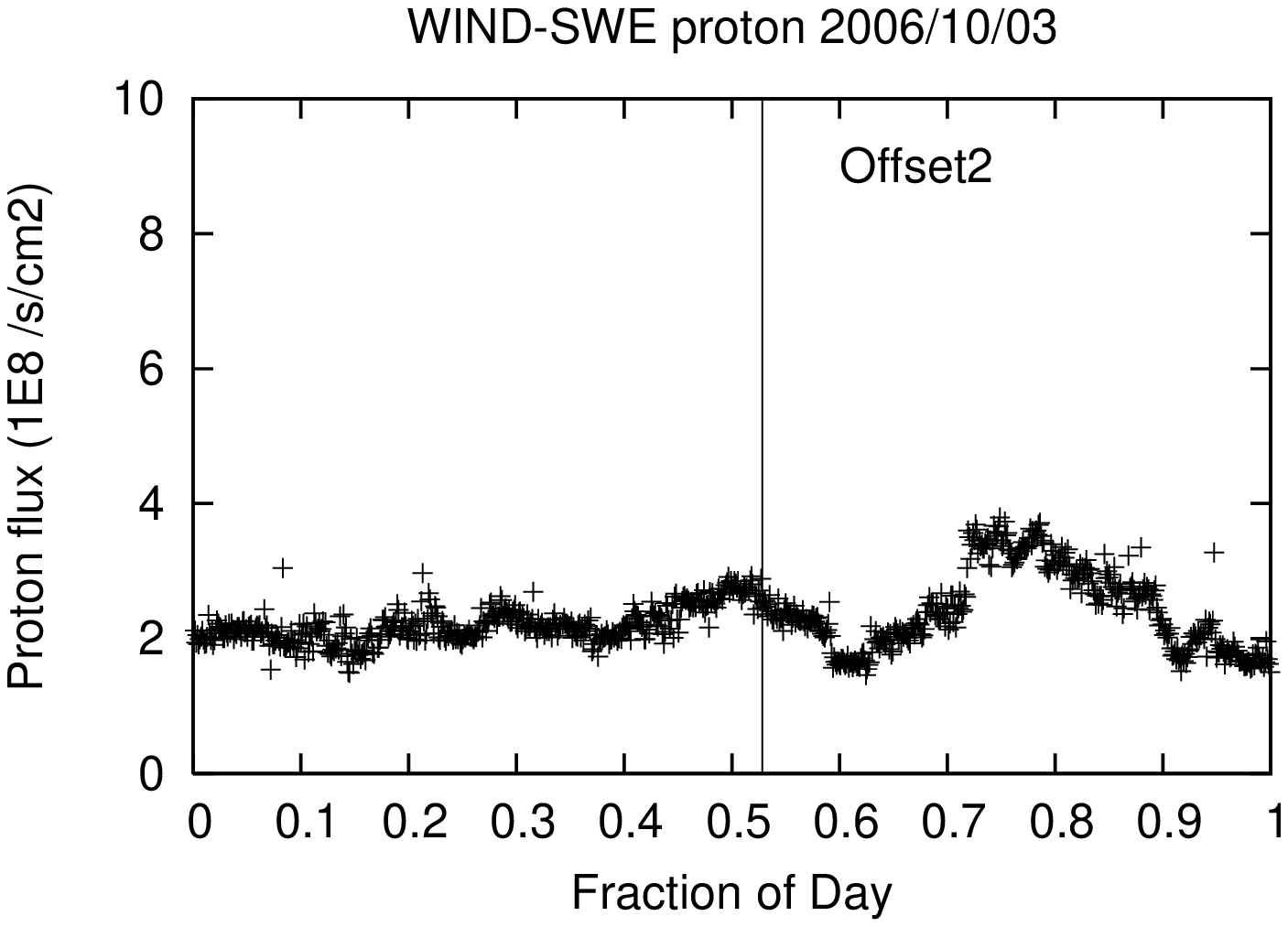} 
\hspace{0.2cm}
\vspace{0.2cm}
   \includegraphics[width=0.2\textwidth,angle=-90,clip]{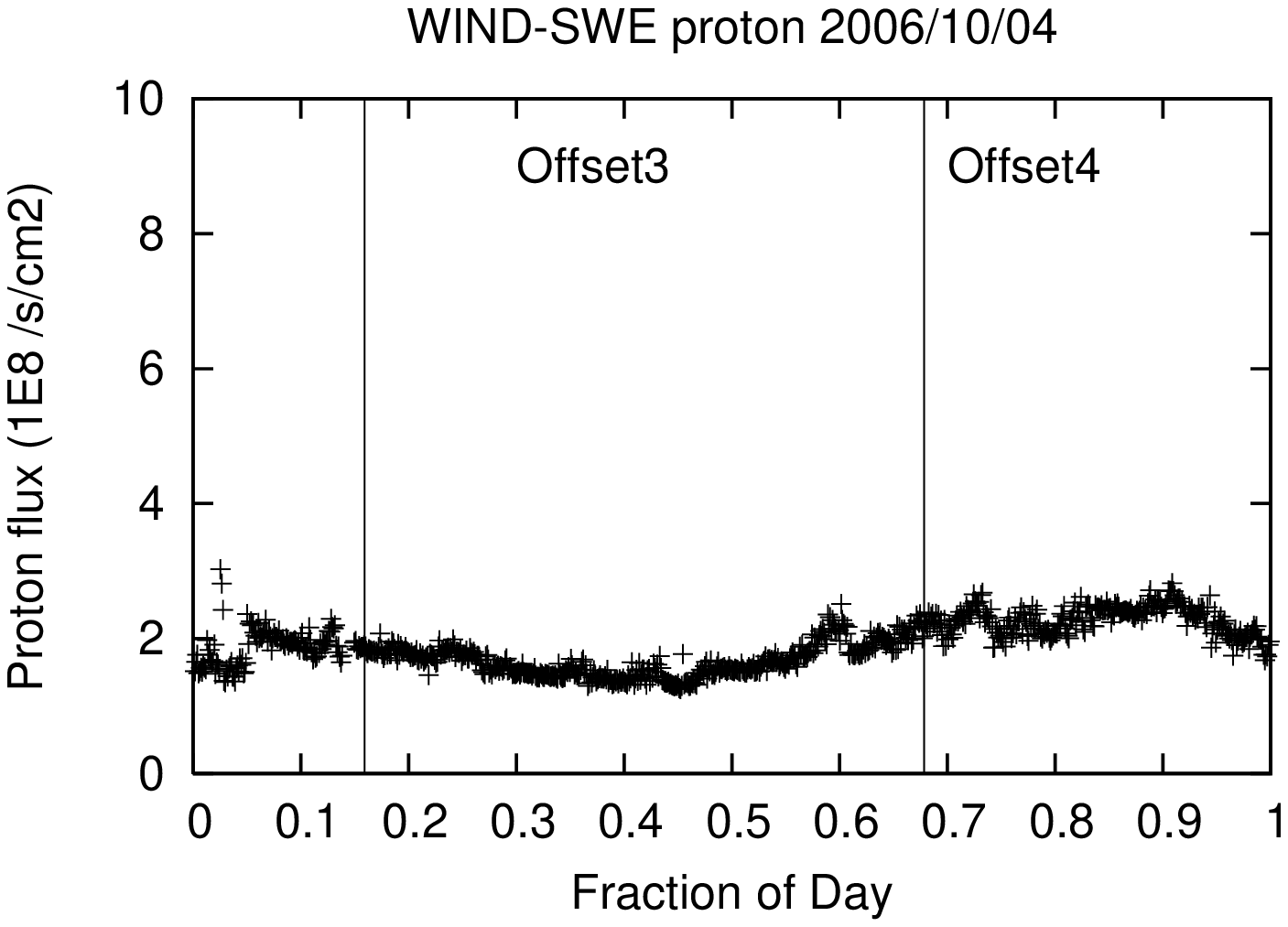} 
\hspace{0.2cm}
\vspace{0.2cm}
   \includegraphics[width=0.2\textwidth,angle=-90,clip]{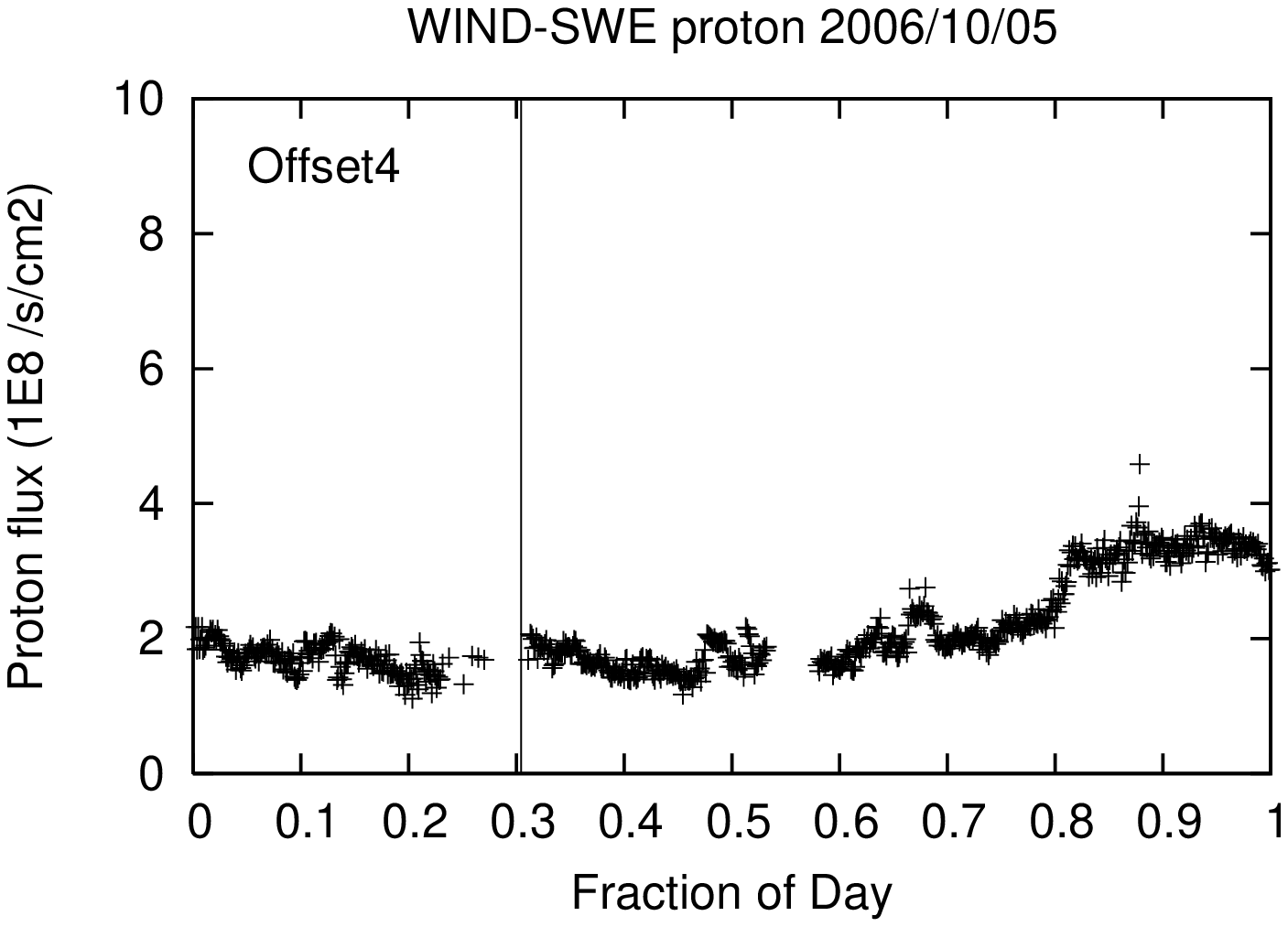} 
  \end{center}
  \caption{
Proton flux (speed times density) measured by the WIND-SWE when XMM-Newton and 
Suzaku observed Abell~2199. 
}
\label{fig:wind-lc}
\end{figure}

figure~\ref{fig:wind-lc} is proton flux of WIND-SWE\footnote{http://web.mit.edu/space/www/wind.html}, when XMM-Newton and Suzaku observed Abell~2199. 
These plots are created by multiplying proton speed and
proton density of which data are public in the WIND-SWE web site. 

\section{Fitting results of the seven annular regions
\label{appen:fit-result}}

%
%


\newpage



\begin{figure}
  \begin{center}
   \includegraphics[width=0.25\textwidth,angle=-90,clip]{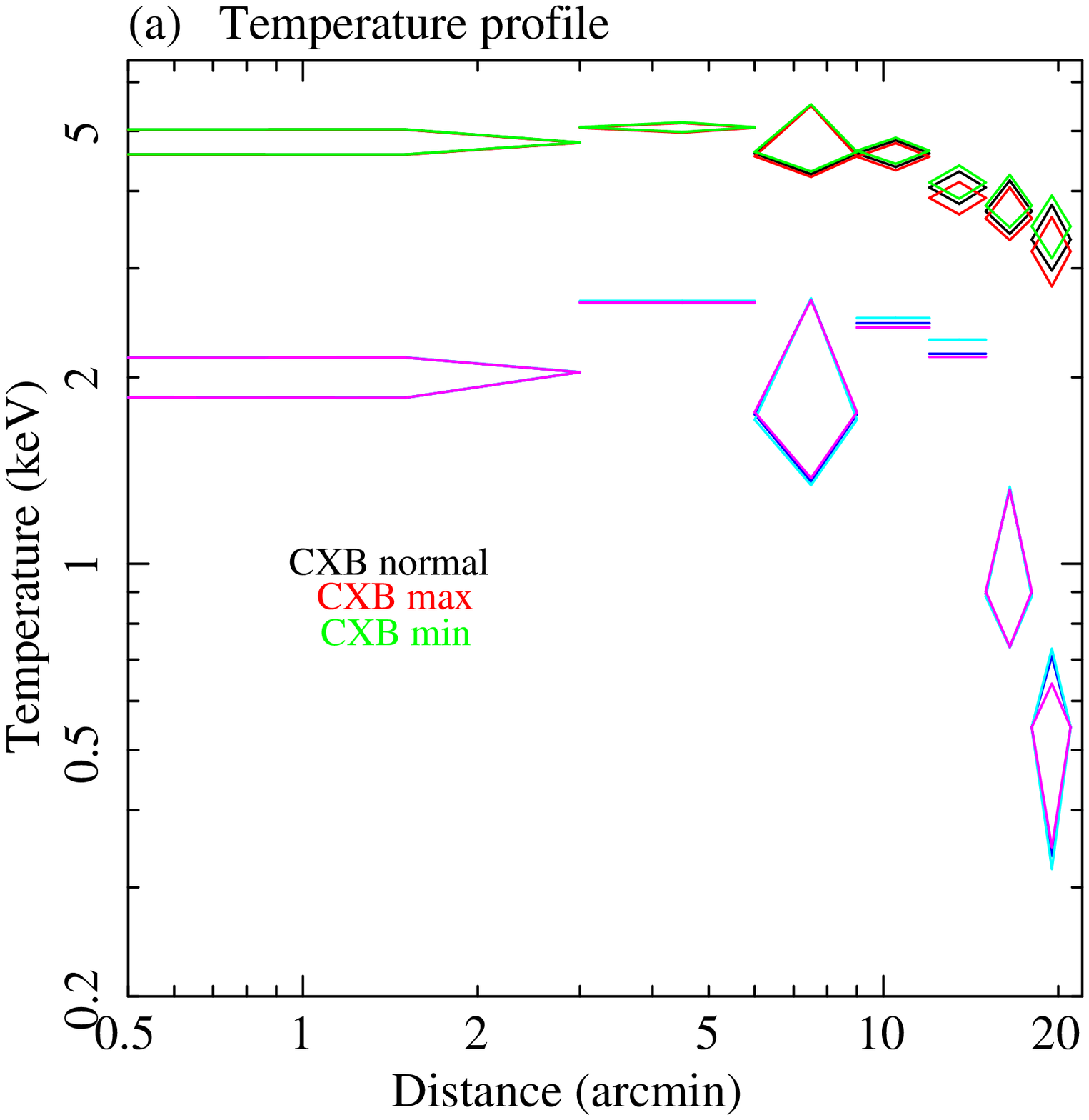} 
\hspace{0.3cm}
\vspace{0.3cm}
   \includegraphics[width=0.25\textwidth,angle=-90,clip]{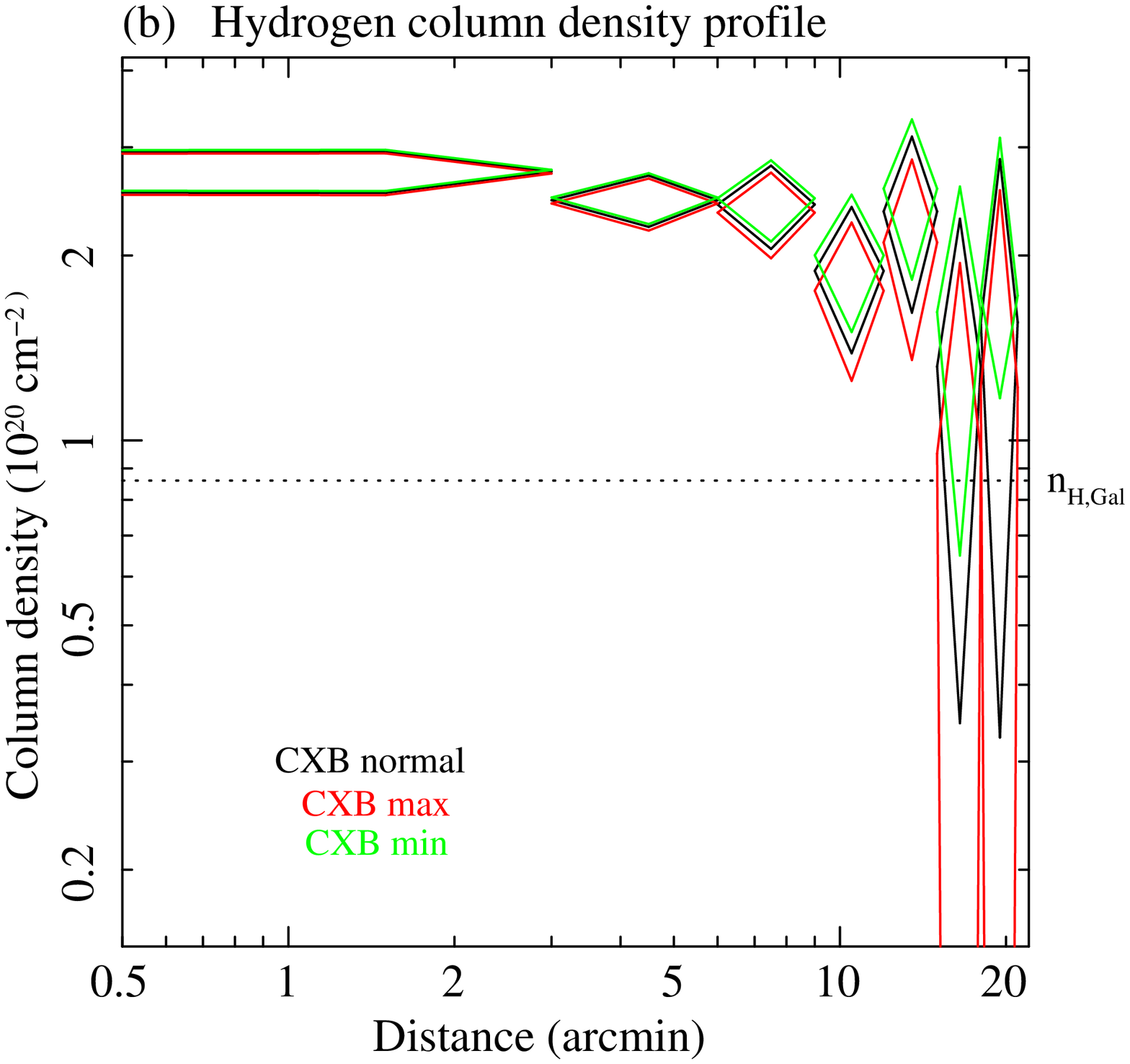} 
\hspace{0.3cm}
   \includegraphics[width=0.25\textwidth,angle=-90,clip]{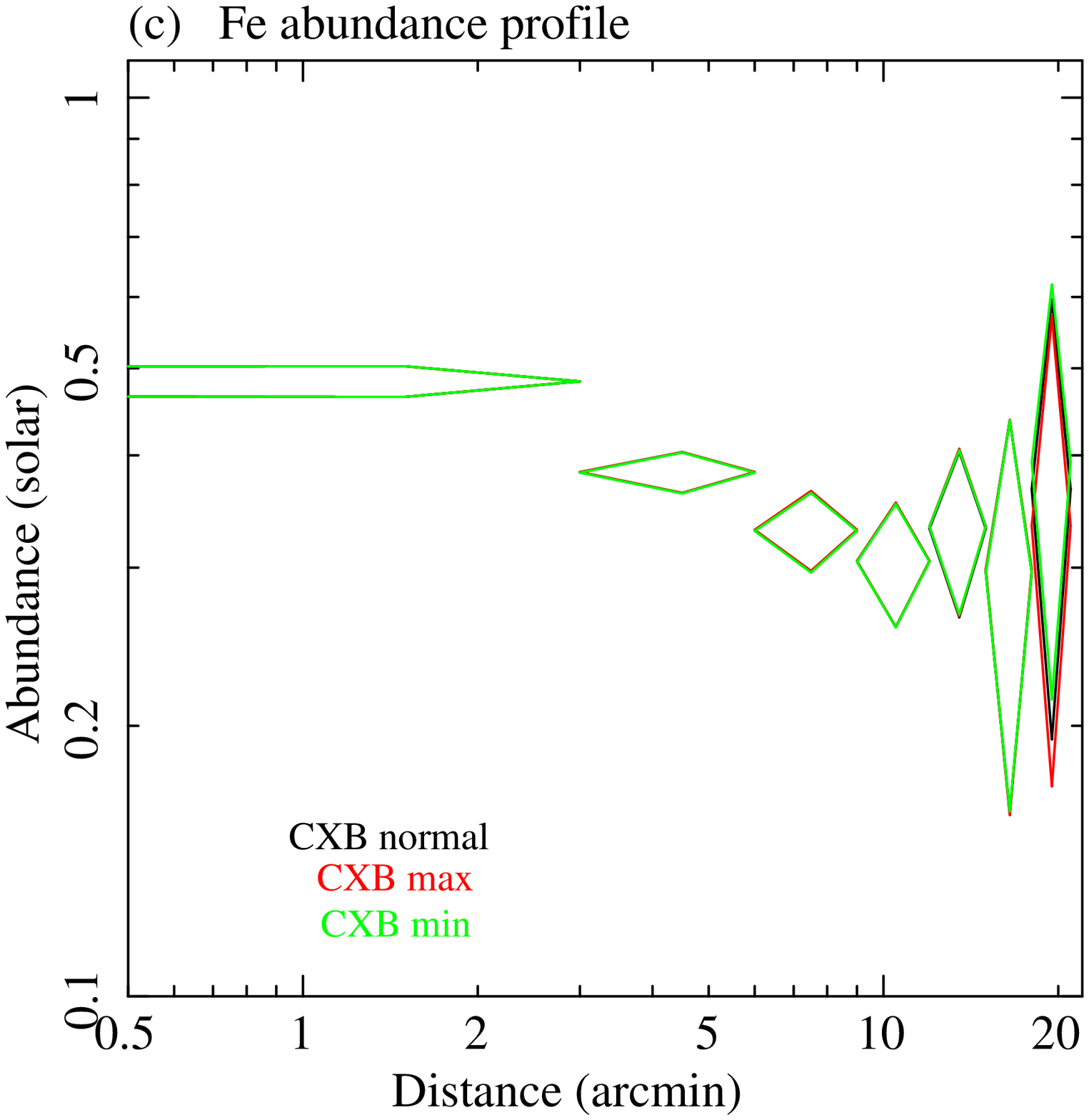} 
\hspace{0.3cm}
   \includegraphics[width=0.25\textwidth,angle=-90,clip]{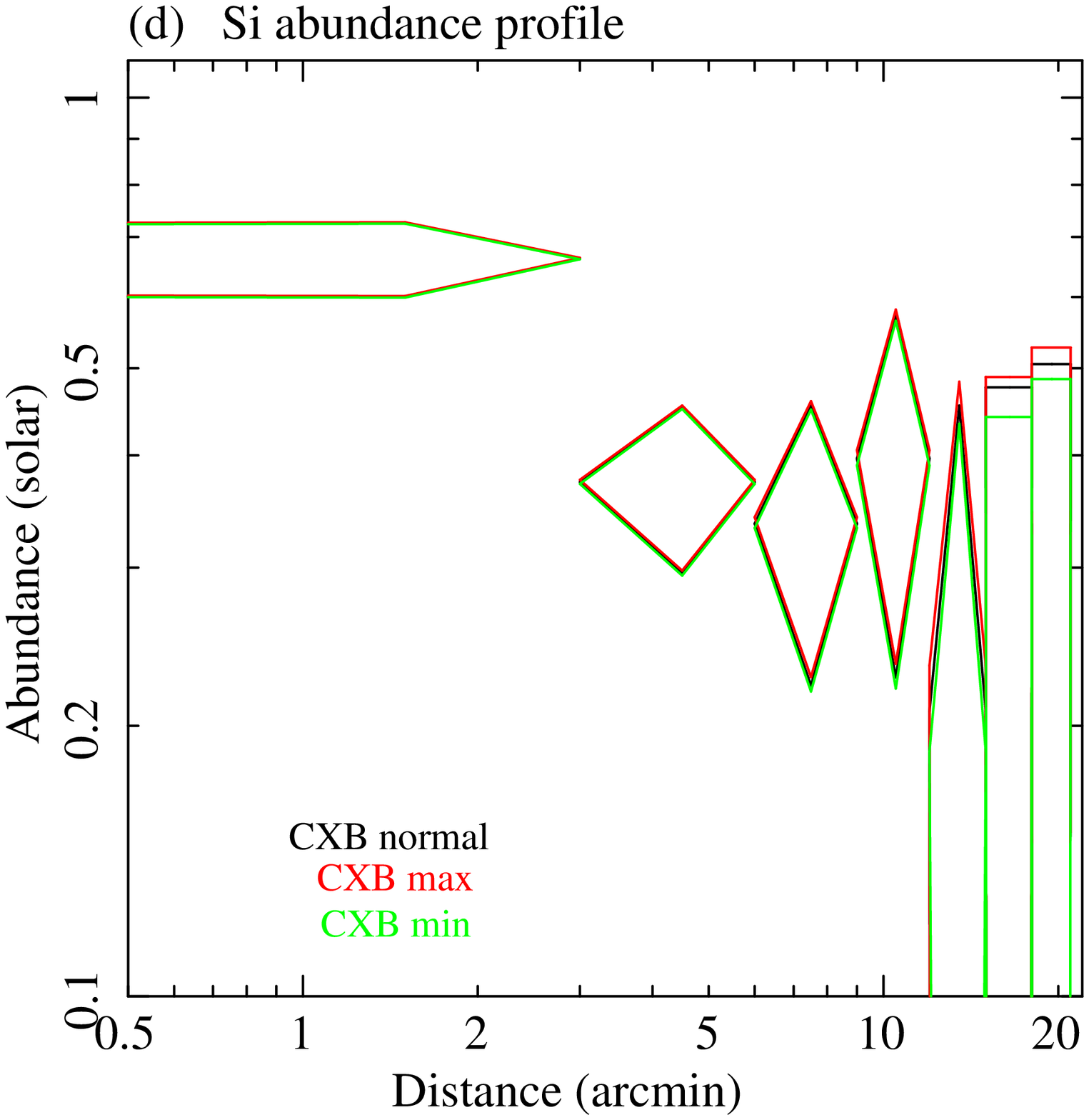} 
\hspace{0.3cm}
   \includegraphics[width=0.25\textwidth,angle=-90,clip]{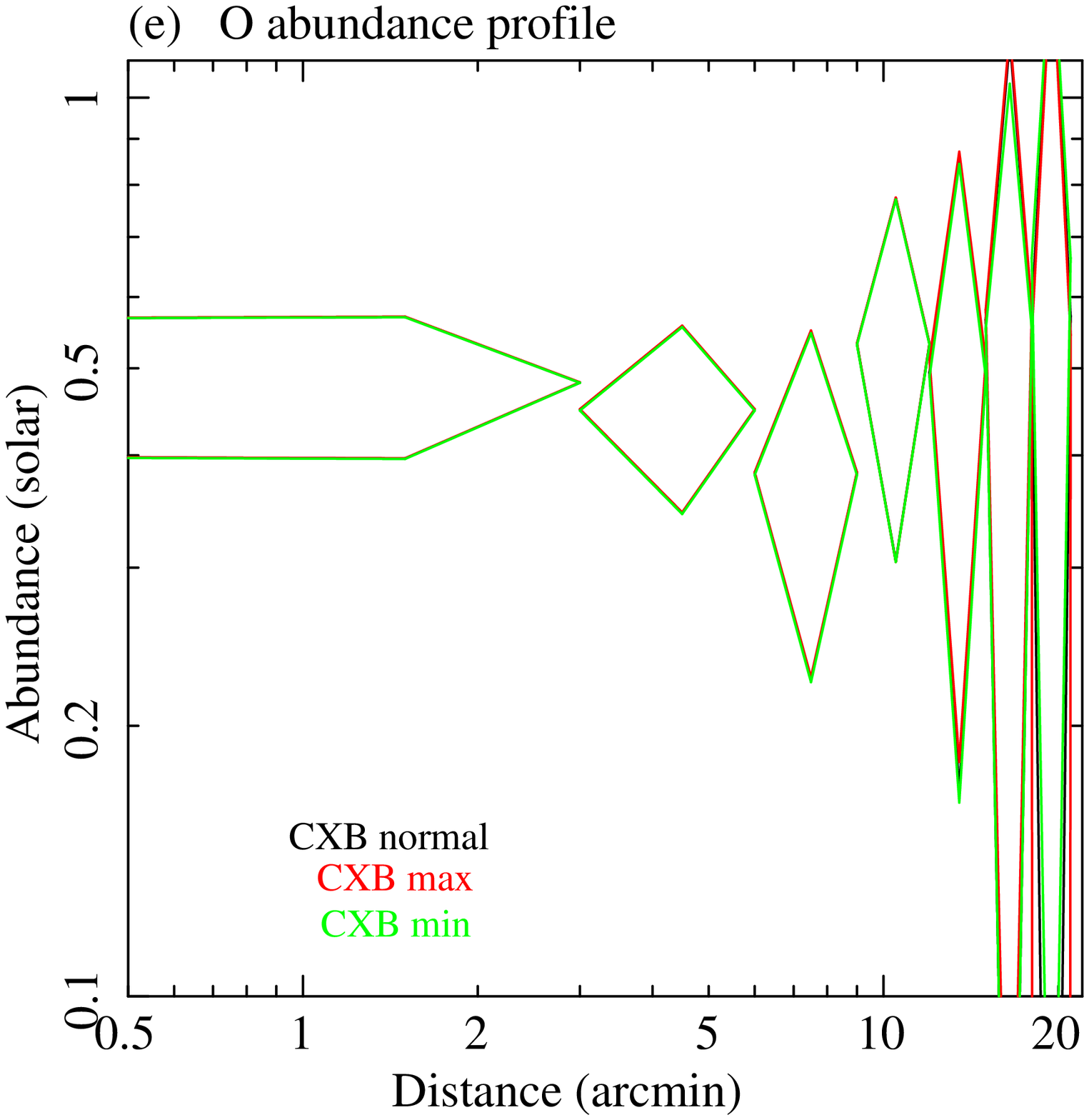} 
\hspace{0.3cm}
   \includegraphics[width=0.25\textwidth,angle=-90,clip]{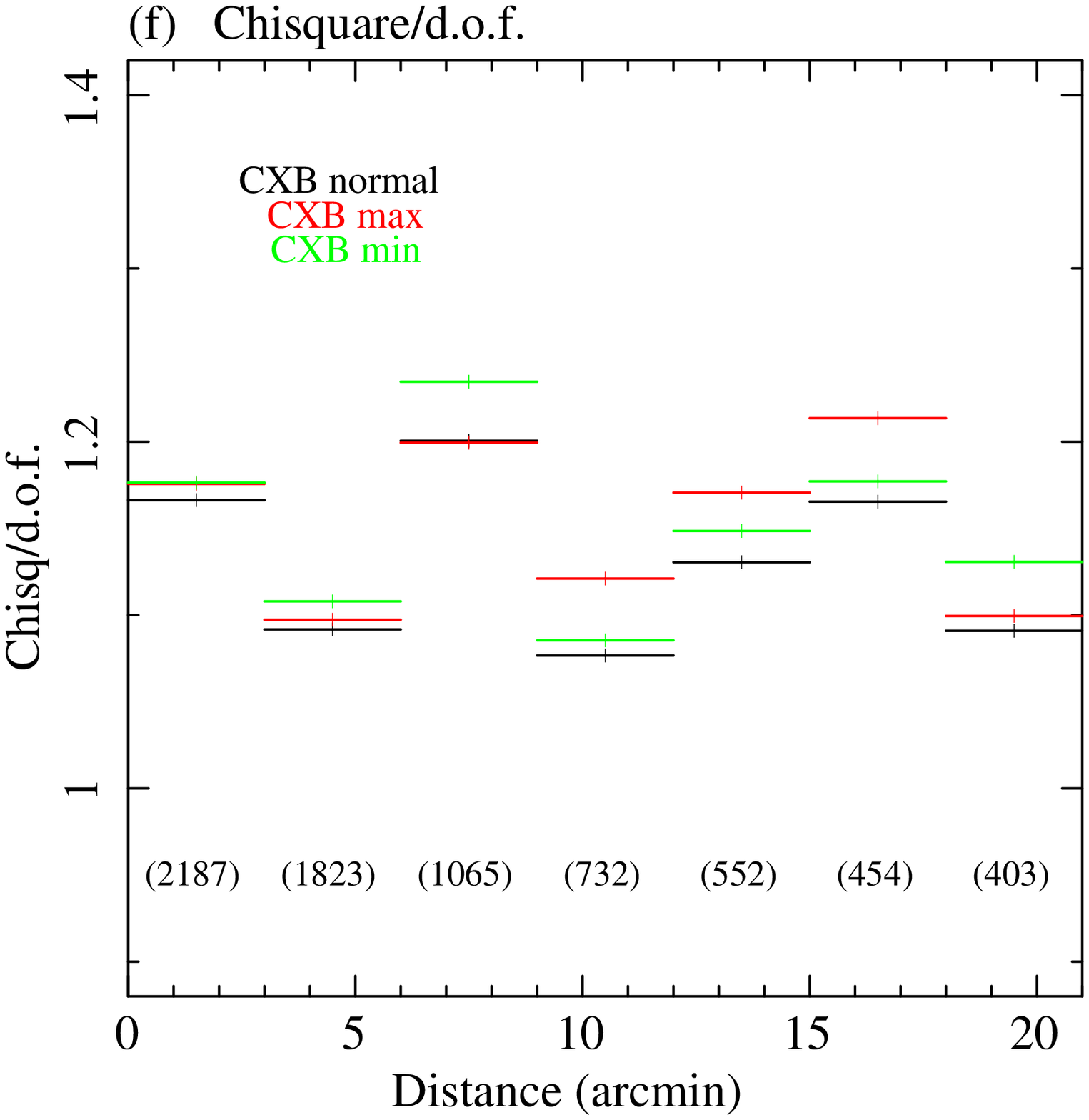} 
  \end{center} 
  \caption{The same as figure~\ref{fig:vapec-prof}, but for the two temperature {\it vapec} fits.
Blue data points in panel (a) indicate the lower-temperature component. In the $3'$--$6'$, 
$9'$--$12'$, and $12'$--$15'$ regions, they were fixed at 0.5 times the corresponding 
best-fit higher 
temperature because they were not constrained. 
Magenta and cyan data points are the lower-temperature component when the CXB level is set
maximam and minimum, respectively. 
See the electronic version of the paper for colour figures. 
}
\label{fig:2vapec-prof}
\end{figure}


\begin{figure}
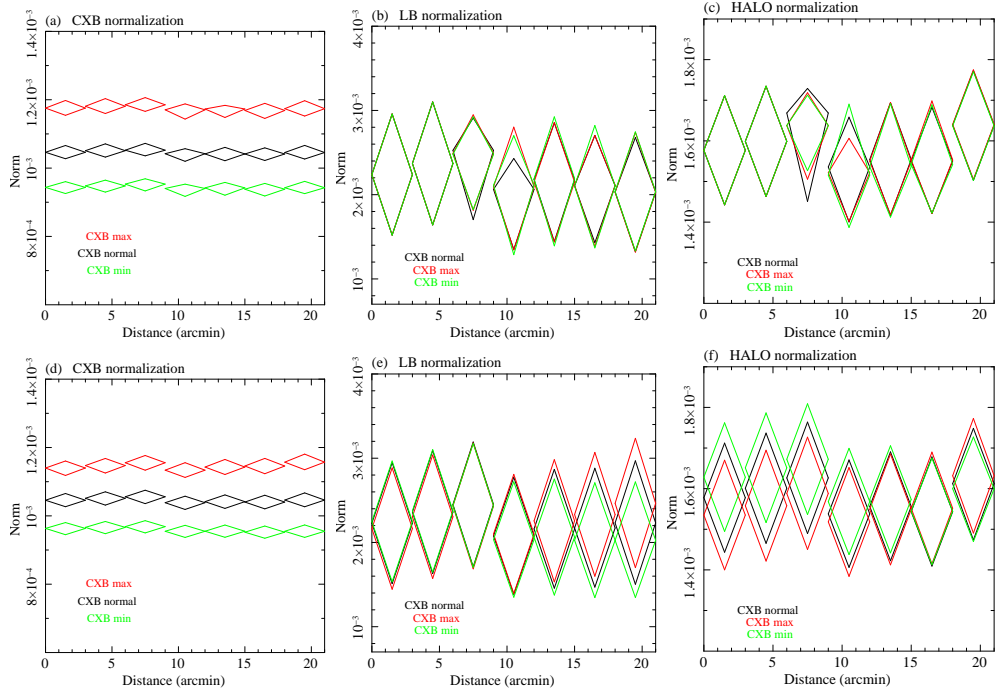

  \begin{center}
   \includegraphics[width=0.27\textwidth,angle=-90,clip]{./figure10a.eps} 
   \includegraphics[width=0.27\textwidth,angle=-90,clip]{./figure10b.eps} 
   \includegraphics[width=0.27\textwidth,angle=-90,clip]{./figure10c.eps} 
   \includegraphics[width=0.27\textwidth,angle=-90,clip]{./figure10d.eps} 
   \includegraphics[width=0.27\textwidth,angle=-90,clip]{./figure10e.eps} 
   \includegraphics[width=0.27\textwidth,angle=-90,clip]{./figure10f.eps} 
  \end{center}
  \caption{
(a) normalization of the CXB determined by the single-temperature {\it vapec}
model (black). Those when the CXB level is set maximum and minimum are also shown
in red and green, respectively. 
(b) The same as panel (a), but for normalization of the LB.
(c) The same as panel (a), but for normalization of the Halo.
(d) The same as panel (a), but for the the two-temperature {\it vapec} model.
(e) The same as panel (b), but for the the two-temperature {\it vapec} model.
(f) The same as panel (c), but for the the two-temperature {\it vapec} model.
See the electronic version of the paper for colour figures. 
}

\label{fig:bgd-norm}
\end{figure}

\clearpage

\begin{table}
  \caption{ Suzaku Observation log of Abell 2199.}
\label{table:observation}
  \begin{center}
    \begin{tabular}{lcccc}
\hline 
\hline 
Position (Obs. ID) & Start$^{*}$ &  End$^{*}$ & XIS Exposure$^{\dagger}$ & PIN exposure$^{\dagger}$\\
\hline 
Center (801056010) & 2006/Oct/01 07:08:00 & 2006/Oct/01 21:31:18 & 18.5 & 19.3\\
Offset1 (801057010)& 2006/Oct/01 21:32:15 & 2006/Oct/02 10:43:19 & 23.5 & 23.7\\
Offset2 (801058010)& 2006/Oct/03 12:41:00 & 2006/Oct/04 03:47:19 & 19.0 & 10.8\\
Offset3 (801059010)& 2006/Oct/04 03:48:52 & 2006/Oct/04 16:17:19 & 18.9 & --\\
Offset4 (801060010)& 2006/Oct/04 16:19:16 & 2006/Oct/05 07:18:24 & 23.5 & 15.3\\
\hline 
    \end{tabular} 
  \end{center}
$^{*}$: {Time is shown in UT.} \\
$^{\dagger}$: {Effective exposure in units of ks, obtained after the event selection.}
\end{table}
\begin{table}
  \caption{HXD-PIN HV operations during the observation$^{*}$.}
\label{table:hv-ope}
  \begin{center}
    \begin{tabular}{lcccc}
\hline 
\hline 
Position & HVP0 & HVP1 & HVP2 & HVP3 \\
\hline 
Center  & 400 & 500 & 500 & 500 \\
Offset1 & 400 & 500 & 500 & 500 \\
Offset2 & 400 & 500 $\to$ 0 & 500 $\to$ 0 & 500 $\to$ 0 \\ 
Offset3 & 400 & 0   & 0   & 0   \\
Offset4 & 400 & 0 $\to$ 400 & 0 $\to$ 500 & 0 $\to$ 500 \\
\hline 
    \end{tabular}
  \end{center}
$^{*}$: {The numbers indicate high voltages in units of volts.} \\
\end{table}

\begin{table}
 \caption{Results of model fittings to XIS spectra of the HLD observation.}
\label{table:fit-bgd}
   \begin{center}
    \begin{tabular}{ccccc}
\hline 
\hline 
$N_{\rm H}$$^{*}$ & $\Gamma$(CXB) & $S_{\rm x}$(CXB)$^{\dagger}$ & $S_{\rm x}$(GFE)$^{\ddagger}$ & $\chi^2/$dof (prob.$^{\S}$) \\
\hline 
$1.02 \times 10^{20}$ (fix)& 1.412 (fix)& $6.35^{+0.13}_{-0.13} \times 10^{-8}$& $1.96^{+0.17}_{-0.17} \times 10^{-8}$ &  315.1/285 (0.11)\\
\hline 
\end{tabular}
\end{center}
$^{*}$: {Hydrogen column density in units of cm$^{-2}$.} \\
$^{\dagger}$: {The 2.0--10.0 keV CXB surface brightness in units of erg s$^{-1}$ cm$^{-2}$ sr$^{-1}$.}\\
$^{\ddagger}$: {The 0.3--1.0 keV GFE surface brightness in units of erg s$^{-1}$ cm$^{-2}$ sr$^{-1}$.} \\
$^{\S}$: {Null hypothesis probability.}
\end{table}
\begin{table}
 \caption{RASS averaged count rates around Abell~2199 and the HLD field.}
\label{table:rass-cnt}
   \begin{center}
    \begin{tabular}{lcc}
\hline 
\hline 
Band$^{*}$ & $1.0^{\circ}$--$2.0^{\circ}$$^{\dagger}$ & HLD$^{\ddagger}$  \\
\hline 
1/4 keV  & $1454.7 \pm 8.2$ & $1745.1 \pm 17.4$ \\
3/4 keV  & $144.0 \pm 2.8$   & $143.1 \pm 5.9$   \\
1.5 keV  & $133.4 \pm 2.7$   & $139.4 \pm 4.6$   \\
\hline 
\end{tabular}
\end{center}
$^{*}$: {RASS 1/4 keV, 3/4 keV, and 1.5 keV bands correspond to
0.12-0.284 keV, 0.47-1.21 keV, and 0.76-2.04 keV, respectively.} \\
$^{\dagger}$: {RASS count rate in an annulus around Abell~2199, in units of 10$^{-6}$ counts s$^{-1}$ arcmin$^{-2}$.}\\
$^{\ddagger}$: {RASS count rate averaged over the HLD field 
($1^{\circ}$ radius) in units of 10$^{-6}$ counts s$^{-1}$ arcmin$^{-2}$.} \\
\end{table}

\begin{table}
 \caption{Observations used in the XIS annular spectrum fitting.}
\label{table:fit-reg}
   \begin{center}
    \begin{tabular}{ll}
\hline 
\hline 
Region & Observation \\
\hline 
$0'$--$3'$ & Center, Offset1 \\
$3'$--$6'$ & Center, Offset1, Offset2 \\
$6'$--$9'$ & Center, Offset1, Offset2 \\
$9'$--$12'$ & Center, Offset1, Offset2, Offset4 \\
$12'$--$15'$ &Offset1, Offset2, Offset3, Offset4 \\
$15'$--$18'$ &Offset1, Offset2, Offset3, Offset4 \\
$18'$--*  & Offset3, Offset4 \\
\hline 
\end{tabular}
\end{center}
\end{table}
\begin{table}
  \caption{Upper limit on the soft excess.}
\label{table:warm-upper}
  \begin{center}
    \begin{tabular}{lccccccc}
\hline 
\hline 
Region & $0'$--$3'$ & $3'$--$6'$ & $6'$--$9'$ & $9'$--$12'$ & $12'$--$15'$ & $15'$--$18'$ & $18'$--$*$\\
\hline 
Warm gas$^{*}$ & 7.5  & 7.3  & 4.0  & 6.5   & 3.7    & 6.0    & 5.3  \\
Power law$^{\dagger}$ &  3.8 & 1.2  & 0.7  & 0.5   & 0.3  & 0.4 & 0.3 \\
\hline 
    \end{tabular}
  \end{center}
$^{*}$: {Surface brightness of 0.2 keV gas, expressed as emission measure per unit solid angle in units of $10^{62}$ cm$^{-3}$ arcmin$^{-2}$} \\
$^{\dagger}$: {0.2--10 keV luminiosity per unit solid angle of a power law with photon index of 2.0, in units of $10^{41}$ erg s$^{-1}$ arcmin$^{-2}$.}
\end{table}

\begin{longtable}{ccccccc}
\label{table:fit-apec}
\multicolumn{7}{l}{Fitting result of the single-temperature {\it vapec} model for the seven annular regions.} \\
\hline
\hline
Region & $kT^{*}$ & $n_{\rm H}^{\dagger}$ & $Z_{\rm Fe}^{\ddagger}$ &  $Z_{\rm Si}^{\ddagger}$ & $Z_{\rm O}^{\ddagger}$ & $\chi^2$/D.O.F. \\
\hline
\endfirsthead 
\multicolumn{7}{l}{Fitting result. --continued} \\
\hline
\hline
Region & $kT^{*}$ & $n_{\rm H}^{\dagger}$ & $Z_{\rm Fe}^{ddagger}$ &  $Z_{\rm Si}^{\ddagger}$ & $Z_{\rm O}^{\ddagger}$ & $\chi^2$/D.O.F. \\
\hline
\endhead 
\hline
\endfoot 
 %
\multicolumn{7}{l}{$^{*}$: Temperature in units of keV. } \\
\multicolumn{7}{l}{$^{\dagger}$: Hydrogen column density in units of $10^{20}$ cm$^{-2}$.} \\
\multicolumn{7}{l}{$^{\ddagger}$: Metal abundance in solar unit of \citet{anders-1989}.} \\
\endlastfoot 
\multicolumn{7}{c}{CXB level is default} \\
\hline
$0' $--$ 3' $ & $3.91^{+0.04}_{-0.04}$ & $2.71^{+0.21}_{-0.20}$ & $0.52^{+0.02}_{-0.02}$ & $0.66^{+0.07}_{-0.07}$ & $0.73^{+0.10}_{-0.09}$ & $2687.5/2189$ \\ 
$3' $--$ 6' $ & $4.26^{+0.05}_{-0.05}$ & $2.43^{+0.24}_{-0.23}$ & $0.39^{+0.02}_{-0.02}$ & $0.37^{+0.08}_{-0.08}$ & $0.55^{+0.12}_{-0.12}$ & $2014.8/1823$ \\ 
$6' $--$ 9' $ & $4.10^{+0.06}_{-0.09}$ & $2.25^{+0.50}_{-0.26}$ & $0.36^{+0.03}_{-0.03}$ & $0.34^{+0.11}_{-0.11}$ & $0.55^{+0.18}_{-0.12}$ & $1295.1/1067$ \\ 
$9' $--$ 12'$ & $3.90^{+0.12}_{-0.10}$ & $1.88^{+0.46}_{-0.57}$ & $0.33^{+0.05}_{-0.04}$ & $0.38^{+0.19}_{-0.17}$ & $0.65^{+0.27}_{-0.23}$ & $790.5/732  $ \\ 
$12'$--$15' $ & $3.57^{+0.13}_{-0.21}$ & $2.06^{+1.03}_{-0.56}$ & $0.38^{+0.07}_{-0.08}$ & $0.20^{+0.27}_{-0.20}$ & $0.63^{+0.44}_{-0.26}$ & $625.1/552  $ \\ 
$15'$--$18' $ & $3.61^{+0.15}_{-0.38}$ & $0.50^{+1.41}_{-0.50}$ & $0.30^{+0.11}_{-0.09}$ & $<0.24             $ & $0.72^{+0.56}_{-0.47}$ & $539.3/456  $ \\ 
$18'$--$21' $ & $3.10^{+0.36}_{-0.34}$ & $0.80^{+0.95}_{-0.80}$ & $0.27^{+0.19}_{-0.14}$ & $<0.33             $ & $0.57^{+0.87}_{-0.53}$ & $448.5/406  $ \\ 
\hline
\multicolumn{7}{c}{CXB level is minimum} \\
\hline
$0' $--$ 3' $ & $3.91^{+0.04}_{-0.04}$ & $2.71^{+0.21}_{-0.20}$ & $0.52^{+0.02}_{-0.02}$ & $0.66^{+0.07}_{-0.07}$ & $0.73^{+0.10}_{-0.09}$ & $2688.0/2189$ \\
$3' $--$ 6' $ & $4.26^{+0.05}_{-0.05}$ & $2.43^{+0.24}_{-0.23}$ & $0.39^{+0.02}_{-0.02}$ & $0.37^{+0.08}_{-0.08}$ & $0.55^{+0.12}_{-0.12}$ & $2015.6/1823$ \\
$6' $--$ 9' $ & $4.10^{+0.07}_{-0.08}$ & $2.34^{+0.39}_{-0.34}$ & $0.36^{+0.03}_{-0.03}$ & $0.33^{+0.11}_{-0.11}$ & $0.55^{+0.16}_{-0.13}$ & $1295.6/1067$ \\
$9' $--$ 12'$ & $3.94^{+0.11}_{-0.11}$ & $1.82^{+0.50}_{-0.51}$ & $0.33^{+0.05}_{-0.05}$ & $0.39^{+0.18}_{-0.17}$ & $0.65^{+0.26}_{-0.22}$ & $790.6/732  $ \\
$12'$--$15' $ & $3.56^{+0.18}_{-0.14}$ & $2.38^{+0.64}_{-0.88}$ & $0.37^{+0.08}_{-0.06}$ & $0.19^{+0.27}_{-0.19}$ & $0.65^{+0.40}_{-0.19}$ & $624.8/552  $ \\
$15'$--$18' $ & $3.68^{+0.19}_{-0.37}$ & $0.30^{+1.53}_{-0.30}$ & $0.30^{+0.12}_{-0.09}$ & $<0.24             $ & $0.69^{+0.60}_{-0.47}$ & $539.3/456  $ \\
$18'$--$21' $ & $3.27^{+0.38}_{-0.33}$ & $0.71^{+0.95}_{-0.71}$ & $0.30^{+0.19}_{-0.14}$ & $<0.33             $ & $0.64^{+0.87}_{-0.58}$ & $446.1/406  $ \\
\hline
\multicolumn{7}{c}{CXB level is maximum} \\
\hline
$0' $--$ 3' $ & $3.91^{+0.04}_{-0.04}$ & $2.71^{+0.21}_{-0.20}$ & $0.52^{+0.02}_{-0.02}$ & $0.66^{+0.07}_{-0.07}$ & $0.73^{+0.10}_{-0.09}$ & $2686.9/2189$ \\
$3' $--$ 6' $ & $4.26^{+0.05}_{-0.05}$ & $2.43^{+0.24}_{-0.23}$ & $0.39^{+0.02}_{-0.02}$ & $0.37^{+0.08}_{-0.08}$ & $0.55^{+0.12}_{-0.12}$ & $2013.9/1823$ \\
$6' $--$ 9' $ & $4.07^{+0.07}_{-0.07}$ & $2.38^{+0.36}_{-0.36}$ & $0.36^{+0.03}_{-0.03}$ & $0.33^{+0.11}_{-0.11}$ & $0.55^{+0.16}_{-0.14}$ & $1293.9/1067$ \\
$9' $--$ 12'$ & $3.87^{+0.11}_{-0.11}$ & $1.83^{+0.52}_{-0.49}$ & $0.33^{+0.05}_{-0.05}$ & $0.40^{+0.17}_{-0.09}$ & $0.65^{+0.27}_{-0.21}$ & $790.3/732  $ \\
$12'$--$15' $ & $3.48^{+0.13}_{-0.19}$ & $2.11^{+1.03}_{-0.58}$ & $0.37^{+0.08}_{-0.08}$ & $0.20^{+0.28}_{-0.20}$ & $0.64^{+0.42}_{-0.27}$ & $625.6/552  $ \\
$15'$--$18' $ & $3.45^{+0.19}_{-0.35}$ & $0.45^{+1.59}_{-0.45}$ & $0.28^{+0.13}_{-0.06}$ & $<0.26             $ & $0.72^{+0.47}_{-0.50}$ & $539.4/456  $ \\
$18'$--$21' $ & $2.89^{+0.37}_{-0.38}$ & $0.88^{+1.04}_{-0.88}$ & $0.25^{+0.18}_{-0.13}$ & $<0.32             $ & $0.53^{+0.80}_{-0.53}$ & $452.2/406  $ \\
\hline
\end{longtable}

\clearpage

\begin{longtable}{ccccccc}
\label{table:fit-2apec}
\multicolumn{7}{l}{Fitting result of the two-temperature {\it vapec} model for the seven annular regions.} \\
\hline
\hline
Region & $kT^{*}$ & $n_{\rm H}^{\dagger}$ & $Z_{\rm Fe}^{\ddagger}$ &  $Z_{\rm Si}^{\ddagger}$ & $Z_{\rm O}^{\ddagger}$ & $\chi^2$/D.O.F. \\
\hline
\endfirsthead 
\multicolumn{7}{l}{Fitting result. --continued} \\
\hline
\hline
Region & $kT^{*}$ & $n_{\rm H}^{\dagger}$ & $Z_{\rm Fe}^{ddagger}$ &  $Z_{\rm Si}^{\ddagger}$ & $Z_{\rm O}^{\ddagger}$ & $\chi^2$/D.O.F. \\
\hline
\endhead 
\hline
\endfoot 
 %
\multicolumn{7}{l}{$^{*}$: Temperatures of (higher component) / (lower component) in units of keV. } \\
\multicolumn{7}{l}{$^{\dagger}$: Hydrogen column density in units of $10^{20}$ cm$^{-2}$.} \\
\multicolumn{7}{l}{$^{\ddagger}$: Metal abundance in solar unit of \citet{anders-1989}.} \\
\endlastfoot 
\multicolumn{7}{c}{CXB level is default} \\
\hline
$0' $--$ 3' $ & $4.79^{+0.24}_{-0.21} / 2.04^{+0.11}_{-0.19}$  & $2.74^{+0.21}_{-0.21}$ & $0.48^{+0.02}_{-0.02}$ & $0.66^{+0.06}_{-0.06}$ &  $0.48^{+0.09}_{-0.09}$ & $2550.8/2187$ \\
$3' $--$ 6' $ & $5.07^{+0.09}_{-0.09} / 2.65$ (fix)	    & $2.46^{+0.24}_{-0.23}$ & $0.38^{+0.02}_{-0.02}$  & $0.37^{+0.08}_{-0.08}$ &  $0.45^{+0.11}_{-0.10}$ & $1990.2/1823 $ \\
$6' $--$ 9' $ & $4.60^{+0.91}_{-0.34} / 1.74^{+0.94}_{-0.39}$  & $2.42^{+0.38}_{-0.37}$ & $0.33^{+0.03}_{-0.03}$ & $0.34^{+0.12}_{-0.11}$  & $0.38^{+0.17}_{-0.16}$ & $1278.6/1065 $ \\
$9' $--$ 12'$ & $4.60^{+0.23}_{-0.22} / 2.45$ (fix)	    & $1.89^{+0.51}_{-0.50}$  & $0.30^{+0.05}_{-0.05}$ & $0.40^{+0.18}_{-0.17}$ &  $0.53^{+0.24}_{-0.23}$ & $788.1/732   $ \\
$12'$--$15' $ & $4.05^{+0.25}_{-0.24} / 2.18$ (fix)	    & $2.36^{+0.76}_{-0.75}$  & $0.33^{+0.07}_{-0.07}$ & $0.21^{+0.25}_{-0.21}$ &  $0.49^{+0.36}_{-0.32}$ & $624.0/552   $ \\
$15'$--$18' $ & $3.71^{+0.45}_{-0.30} / 0.89^{+0.43}_{-0.16}$  & $1.32^{+0.98}_{-0.97}$  & $0.30^{+0.14}_{-0.14}$ &$<0.48               $  & $0.57^{+0.59}_{-0.54}$ & $529.1/454   $ \\
$18'$--$21' $ & $3.34^{+0.46}_{-0.36} / 0.54^{+0.16}_{-0.21}$  & $1.56^{+1.31}_{-1.23}$  & $0.37^{+0.23}_{-0.17}$ &$<0.51               $  & $0.57^{+0.94}_{-0.56}$ & $439.6/403   $ \\
\hline
\multicolumn{7}{c}{CXB level is minimum} \\
\hline
$0' $--$ 3' $ & $4.79^{+0.24}_{-0.21} / 2.04^{+0.11}_{-0.19}$ &   $2.76^{+0.21}_{-0.21}$ &  $0.48^{+0.02}_{-0.02}$ &  $0.66^{+0.06}_{-0.06}$ & $0.48^{+0.09}_{-0.09}$ & $2572.8/2187$ \\
$3' $--$ 6' $ & $5.07^{+0.09}_{-0.09} / 2.64$ (fix)          &  $2.48^{+0.24}_{-0.24}$ &  $0.38^{+0.02}_{-0.02}$ &  $0.37^{+0.08}_{-0.08}$ & $0.45^{+0.11}_{-0.10}$ & $2019.8/1823$ \\
$6' $--$ 9' $ & $4.63^{+0.89}_{-0.32} / 1.76^{+0.91}_{-0.38}$ &   $2.48^{+0.38}_{-0.37}$ &  $0.33^{+0.03}_{-0.03}$ &  $0.33^{+0.12}_{-0.11}$ & $0.38^{+0.17}_{-0.16}$ & $1314.9/1065$ \\
$9' $--$ 12'$ & $4.65^{+0.23}_{-0.22} / 2.41$ (fix)          &  $2.00^{+0.51}_{-0.50}$ &  $0.30^{+0.05}_{-0.05}$ &  $0.39^{+0.17}_{-0.17}$ & $0.53^{+0.24}_{-0.23}$ & $794.5/732  $ \\
$12'$--$15' $ & $4.13^{+0.27}_{-0.24} / 2.16$ (fix)          &  $2.57^{+0.76}_{-0.74}$ &  $0.33^{+0.07}_{-0.07}$ &  $0.19^{+0.25}_{-0.19}$ & $0.49^{+0.35}_{-0.32}$ & $634.0/552  $ \\
$15'$--$18' $ & $3.79^{+0.46}_{-0.30} / 0.90^{+0.42}_{-0.17}$ &   $1.62^{+0.97}_{-0.97}$ &  $0.30^{+0.14}_{-0.14}$ &  $<0.44             $ & $0.55^{+0.48}_{-0.53}$ & $534.4/454  $ \\
$18'$--$21' $ & $3.51^{+0.42}_{-0.39} / 0.54^{+0.10}_{-0.20}$ &   $1.72^{+1.39}_{-0.55}$ &  $0.39^{+0.23}_{-0.18}$ &  $<0.49             $ & $0.66^{+0.87}_{-0.63}$ & $455.7/403  $ \\
\hline
\multicolumn{7}{c}{CXB level is maximum} \\
\hline
$0' $--$ 3' $ & $4.79^{+0.24}_{-0.21} / 2.04^{+0.11}_{-0.19}$ &  $2.72^{+0.21}_{-0.21}$ & $0.48^{+0.02}_{-0.02}$ & $0.66^{+0.06}_{-0.06}$ &  $0.48^{+0.09}_{-0.09}$ & $2571.3/2187$ \\
$3' $--$ 6' $ & $5.06^{+0.09}_{-0.09} / 2.66$ (fix)         &  $2.43^{+0.24}_{-0.23}$ & $0.38^{+0.02}_{-0.02}$ & $0.38^{+0.08}_{-0.08}$ &  $0.45^{+0.11}_{-0.10}$ & $2000.3/1823$ \\
$6' $--$ 9' $ & $4.55^{+0.94}_{-0.33} / 1.71^{+0.97}_{-0.37}$ &  $2.35^{+0.38}_{-0.37}$ & $0.33^{+0.03}_{-0.03}$ & $0.34^{+0.12}_{-0.11}$ &  $0.38^{+0.17}_{-0.16}$ & $1277.4/1065$ \\
$9' $--$ 12'$ & $4.55^{+0.23}_{-0.23} / 2.49$ (fix)         &  $1.75^{+0.51}_{-0.50}$ & $0.31^{+0.05}_{-0.05}$ & $0.41^{+0.18}_{-0.17}$ &  $0.53^{+0.24}_{-0.23}$ & $820.6/732  $ \\
$12'$--$15' $ & $3.90^{+0.24}_{-0.23} / 2.30$ (fix)         &  $2.10^{+0.77}_{-0.75}$ & $0.33^{+0.07}_{-0.07}$ & $0.23^{+0.25}_{-0.23}$ &  $0.51^{+0.36}_{-0.33}$ & $646.2/552  $ \\
$15'$--$18' $ & $3.61^{+0.44}_{-0.28} / 0.89^{+0.45}_{-0.15}$ &  $0.95^{+0.99}_{-0.95}$ & $0.30^{+0.14}_{-0.14}$ & $<0.49             $ &  $0.58^{+0.60}_{-0.55}$ & $551.0/454  $ \\
$18'$--$21' $ & $3.20^{+0.43}_{-0.39} / 0.54^{+0.18}_{-0.22}$ &  $1.22^{+1.33}_{-1.22}$ & $0.33^{+0.24}_{-0.16}$ & $<0.53             $ &  $0.54^{+0.91}_{-0.54}$ & $443.1/ 403 $ \\
\hline
\end{longtable}

\end{document}